\documentclass[aps,pra,twocolumn,footinbib,showpacs,showkeys]{revtex4-1}

\usepackage{graphicx}
\usepackage{indentfirst}
\usepackage{physics}
\usepackage{braket}
\usepackage{float}
\usepackage{amsmath}
\usepackage{upgreek}
\usepackage{verbatim}
\usepackage{epstopdf}
\usepackage{CJK}
\usepackage{esint}
\usepackage{color}
\usepackage[T1]{fontenc}
\usepackage{subfigure}
\usepackage{amsfonts}
\usepackage{footmisc}
\usepackage{scrextend}
\usepackage{multirow}
\usepackage[hyperfootnotes=false]{hyperref}

\usepackage[english]{babel}
\usepackage{url}
\usepackage{bm}
\usepackage{hyperref}
\definecolor{darkblue}{rgb}{0,0,0.5}
\hypersetup{
colorlinks=true,
linkcolor=black,
filecolor=blue,
citecolor=darkblue,
urlcolor=black,
}

\newenvironment{proof}[1][Proof]{\noindent\textbf{#1.} }{\ \rule{0.5em}{0.5em}}

\newtheorem{innercustomthm}{Theorem}
\newenvironment{customtheorem}[1]
  {\renewcommand\theinnercustomthm{#1}\innercustomthm}
  {\endinnercustomthm}
  
\newtheorem{innercustomthmp}{Corollary}
\newenvironment{customcorollary}[1]
  {\renewcommand\theinnercustomthmp{#1}\innercustomthmp}
  {\endinnercustomthmp}
  
\newtheorem{innercustomthmpp}{Lemma}
\newenvironment{customlemma}[1]
  {\renewcommand\theinnercustomthmpp{#1}\innercustomthmpp}
  {\endinnercustomthmpp}

\newtheorem{innercustomthmppp}{Proposition}
\newenvironment{customproposition}[1]
  {\renewcommand\theinnercustomthmppp{#1}\innercustomthmppp}
  {\endinnercustomthmppp}
  
\newtheorem{innercustomthmpppp}{Remark}
\newenvironment{customremark}[1]
  {\renewcommand\theinnercustomthmpppp{#1}\innercustomthmpppp}
  {\endinnercustomthmpppp}

\newcommand{\calE}{{\cal E}}

\newcommand{\calI}{{\cal I}}

\newcommand{\calT}{{\cal T}}

\newcommand{\1}{^{(1)}}

\def\be{\begin{equation}}
\def\ee{\end{equation}}
\def\ba{\begin{eqnarray}}
\def\ea{\end{eqnarray}}
\usepackage{bm}

\newcommand{\QZ}[1]{{{\textcolor{black}{#1}}}}
\newcommand{\SP}[1]{{{\textcolor{black}{#1}}}}

\begin{document}

\title{Ultimate limits for multiple quantum channel discrimination} %

\author{Quntao Zhuang$^{1,2}$}
\email{zhuangquntao@email.arizona.edu}
\author{Stefano Pirandola$^{3}$}
\affiliation{
$^1$Department of Electrical and Computer Engineering, University of Arizona, Tucson, AZ 85721, USA
\\
$^2$James C. Wyant College of Optical Sciences, University of Arizona, Tucson, AZ 85721, USA
\\
$^3$Department of Computer Science, University of York, York YO10 5GH, UK
}
\date{\today}

\begin{abstract}
Quantum hypothesis testing is a central task in the entire field of quantum information theory. Understanding its ultimate limits will give insight into a wide range of quantum protocols and applications, from sensing to communication. Although the limits of hypothesis testing between quantum states have been completely clarified by the pioneering works of Helstrom in the 70s, the more difficult problem of hypothesis testing with quantum channels, i.e., channel discrimination, is less understood. This is mainly due to the complications coming from the use of input entanglement and the possibility of employing adaptive strategies. In this paper, we establish a lower limit for the ultimate error probability affecting the discrimination of an arbitrary number of quantum channels. We also show that this lower bound is achievable when the channels have certain symmetries. As an example, we apply our results to the problem of channel position finding, where the goal is to identify the location of a target channel among multiple background channels. In this general setting, we find that the use of entanglement offers a great advantage over strategies without entanglement, with non-trivial implications for data readout, target detection and quantum spectroscopy.
\end{abstract}

\maketitle

Hypothesis testing is a fundamental method of statistical inference which plays a central role in both classical and quantum information theory. Since the seminal works by Helstrom~\cite{Helstrom_1976}, quantum hypothesis testing~\cite{Helstrom_1976,Anthony_1998,Chefles_2000,Janos2010} has been greatly advanced for the binary case, namely for the statistical discrimination between two quantum states or two quantum channels. Quantum channel discrimination (QCD)~\cite{KitaevDiamond,Acin_2001,sacchi2005entanglement,wang2006unambiguous,pirandola2018advances} aims at discriminating between different physical processes, modeled as quantum channels and arbitrarily chosen from some known ensemble.
Various protocols have demonstrated the advantages of using entanglement in binary QCD, for example quantum illumination~\cite{tan2008quantum,zhuang2017optimum,zhuang2017NP,zhuang2017quantum,zhang2015} and quantum reading~\cite{Qreading}. It is also known that all resources in any convex resource theory~\cite{takagi2019operational} are useful in binary problems of QCD.

While it is clear that entanglement may give an advantage in some scenarios, the ultimate limit of QCD is far from being understood. The first difficulty results from the fact that solving this limit requires a double optimization, where both input states and output measurements need to be optimized. The second complication comes from the possibility of adaptive strategies, which may strictly outperform non-adaptive ones~\cite{harrow2010adaptive}. So far only special cases have been considered. For unitaries and certain channels, a finite number of probings allow perfect discrimination~\cite{acin2001statistical,duan2009perfect,duan2007entanglement}. For binary discrimination of channels with equal priors, the ultimate adaptive performance can also be found or bounded~\cite{pirandola2019fundamental,pirandola2017ultimate}.

In this paper, we are finally able to address the most general scenario. We establish the ultimate limits for the adaptive discrimination of an arbitrary number of finite-dimensional quantum channels.
More precisely, we provide a general bound to the optimal error probability affecting this general multi-ary discrimination problem, and we also show relevant cases where this bound is achievable. In fact, for a special class of channels with the property of joint teleportation covariance~\cite{pirandola2017ultimate,pirandola2018advances}, our bound is tight and achieved non-adaptively by using maximally-entangled inputs. Furthermore, when the ensemble of channels possesses the geometric uniform symmetry (GUS)~\cite{cariolaro2010theory}, our formulas can be greatly simplified. 

As an application, we study the ultimate minimum error probability for the problem of channel position finding (CPF), where the position of a target channel has to be identified among an array of $m$ cells, with the remaining $m-1$ cells containing copies of a background channel. 
This basic problem has implications for various tasks of quantum sensing \SP{as discussed in Ref.~\cite{CPF2020}}. 
It is here studied considering ensembles of quantum erasure channels (QECs), quantum depolarizing channels (QDCs) and qubit amplitude damping channels (QADCs). In particular, for QDCs, we show that the use of input entanglement strictly outperforms non-entangled strategies. 

\begin{figure*}
\centering
\includegraphics[width=0.9\textwidth]{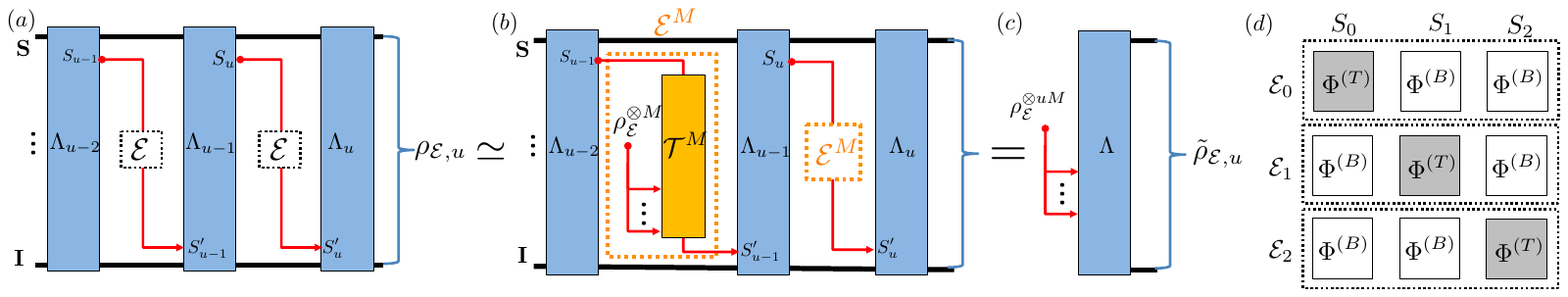}
\caption{Schematics of: (a) A general adaptive protocol. The inputs $\bf S$ and $\bf I$ are quantum registers in an arbitrary state. In the $(k-1)$-th round, a subsystem $S_{k-1}$ probes the channel $\calE$. A quantum operation $\Lambda_{k-1}$ is performed to process the received subsystem $S^\prime_{k-1}$ and prepare the next probe subsystem $S_k$. After $u$ uses, the final decision is made based on the measurement of the output state $\rho_{\calE,u}$; (b) Channel simulation. A general protocol over channel $\calE$ is replaced by a protocol over an approximate channel $\calE^M$, consisting of a teleportation operation ${\calT}^{M}$ applied to $M$ copies of the Choi matrix $\rho_\calE$; (c) Protocol stretching. Starting from the simulated protocol in (b), all the $u$ copies of the resource state $\rho_\calE^{\otimes M}$ are stretched back in time and all the quantum operations (together with the registers $\bf S$ and $\bf I$)  are collapsed into a single trace-preserving quantum operation $\Lambda$ that produces $\tilde{\rho}_{\calE,u}$; (d) Channel position finding with $m=3$ multi-channels $\calE_0,\calE_1,\calE_2$, each acting on three subsystems $S_0,S_1,S_2$. Here $\Phi^{(T)}$ and $\Phi^{(B)}$ represent target and background channels, respectively. \label{fig:schematic_general}}
\end{figure*}

{\em Preliminaries.---} Before addressing QCD, let us summarize the case of state discrimination. The minimum `Helstrom' error probability affecting the discrimination of $m$ states $\{\rho_n\}_{n=0}^{m-1}$ with priors $\{p_n\}_{n=0}^{m-1}$ is given by
\be
P_H\left(\{\rho_n, p_n\}\right)=1-\max_{\sum_n \Pi_n=I}\sum_n p_n {\rm Tr}\left(\rho_n \Pi_n\right),\label{HelstromFIRST}
\ee
where the positive-valued operator measure (POVM) element $\Pi_n$ corresponds to the hypothesis that the state is $\rho_n$. In the binary case with equal priors, it reduces to~\cite{Helstrom_1976}  $P_H=\left(1-\|\rho_1-\rho_2\|/2\right)/2$, where $\|A\|=\tr \sqrt{A^\dagger A}$ is the one-norm. Since evaluating $P_H$ is often challenging, we will resort to various bounds~\cite{PGM1,PGM2,PGM3,Barnum,Bagan,Qiu,Ogawa,montanaro2008lower}~\cite{supp}. To proceed with our study of QCD, we give a continuity bound for $P_H$ as stated in the following lemma (proof in \cite{supp}).
\begin{customlemma}{1}
\label{lemma:helstrom_continuity}
Consider a set of states $\{\rho_n^\prime\}_{n=0}^{m-1}$ close to $\{\rho_n\}_{n=0}^{m-1}$ in the sense that
$
\|\rho_n-\rho_n^\prime\|\le \delta_n$ for $0\le n \le m-1
$.
We lower-bound the Helstrom limit as
\be
P_H\left(\{\rho_n^\prime, p_n\}\right)\ge P_H\left(\{\rho_n, p_n\}\right)-\frac{1}{2}\sum p_n \delta_n.
\ee
\end{customlemma}

{\em Adaptive protocols, simulation and stretching.---} With the continuity bound in hand, we now introduce the most general protocol for QCD and its reduction to state discrimination.
A general $u$-round adaptive protocol for multiple channel discrimination is depicted in Fig.~\ref{fig:schematic_general}(a). The protocol is allowed to access an unknown $d$-dimensional channel $\calE$ for $u$ times, where the unknown channel $\calE$ is fixed but chosen from the ensemble $\{\calE_n, p_n\}_{n=0}^{m-1}$. The unlimited entanglement between all systems involved allows one to push all measurements to the final output $\rho_{\calE,u}$. In each round, a subsystem $S_k, 1\le k \le u$, is sent through the channel $\calE$ and the output $S_k^\prime$ is collected. Our goal is to lower bound the ultimate error probability $P_u$ of the above protocol.

To simplify the structure of the protocol, we employ channel simulation~\cite{quantumPQGA,PLOB,TQCtheory} and protocol stretching~\cite{PLOB}, originally devised for quantum communications. 
As depicted in Fig.~\ref{fig:schematic_general}(b), we consider an approximation $\calE^M$ of the finite-dimensional channel $\calE$ by applying a universal (trace-preserving) teleportation operation $\calT^M$ to $M \ge 1$ copies of the Choi matrix
$
\rho_\calE=\left(\calE\otimes \calI\right) \zeta,
$
where $\zeta:= \sum_{\ell=0}^{d-1} \ket{\ell,\ell}/\sqrt{d}$ is a maximally-entangled state of dimension $d$. In general, $\calT^M$ can be chosen as port-based teleportation (PBT)~\cite{ishizaka2008asymptotic}. The precision of channel simulation is quantified by $\Delta_{\calE,M}:= \|\calE-\calE^M\|_\diamond$ where $\|A\|_\diamond=\sup_{\rho}\|A\otimes \calI \left(\rho\right)\|$ is the diamond norm~\cite{KitaevDiamond,PaulsenBook}. For the simulation of an arbitrary finite-dimensional channel via PBT, we may write~\cite[Lemma~2]{pirandola2019fundamental}
\begin{equation}
\Delta_{\calE,M} \le \delta_{M,d} := 2d(d-1) M^{-1},\label{maxDiamond}
\end{equation}
which is valid for any number of ports $M \ge 1$ and any input dimension $d \ge 2$ for the channel \cite{footnote1}.

The error in the channel simulation propagates to the output of the protocol. Using the triangle inequality, we can bound the trace distance between the output state $\rho_{\calE,u}$ of the actual protocol and the output state $\tilde{\rho}_{\calE,u}$ of the simulated protocol as follows
\be
\|\rho_{\calE,u}-\tilde{\rho}_{\calE,u}\| \le u \Delta_{\calE,M}.\label{errorPROP}
\ee
The final step is protocol stretching\SP{~\cite{PLOB,pirandola2019fundamental}}.
As depicted in Fig.~\ref{fig:schematic_general}(c), this is a re-organization of the simulated protocol into an equivalent block protocol, so that the approximate output state $\tilde{\rho}_{\calE,u}$ is decomposed as  $\tilde{\rho}_{\calE,u}=\Lambda(\rho_\calE^{\otimes u M})$ for a trace-preserving quantum operation $\Lambda$. Combining this with Eq.~(\ref{errorPROP}) we then write
\be
\|\rho_{\calE,u}-\Lambda(\rho_\calE^{\otimes u M})\| \le u \Delta_{\calE,M}.
\label{stretchingEQ}
\ee

{\em Ultimate bounds.---}
Combining Lemma~\ref{lemma:helstrom_continuity} with Eq.~(\ref{stretchingEQ}), we derive the main result of our work (proof in \cite{supp}).
\begin{customtheorem}{1}
\label{theorem:LB}
Consider arbitrary $m \ge 2$ $d-$dimensional quantum channels $\{\calE_n\}_{n=0}^{m-1}$ with prior probabilities $\{p_n\}_{n=0}^{m-1}$. The minimum error probability $P_{u}$ for their $u$-round adaptive discrimination satisfies
\be
P_{u}\ge P_{u,LB}:=P_H\left(\{\rho_{\calE_n}^{\otimes uM}, p_n\}\right)-u\overline{\Delta}_M/2,
\label{bound2}
\ee
where the average simulation error $\overline{\Delta}_M=\sum_n p_n \Delta_{\calE_n,M}$ can be replaced by the uniform error $\delta_{M,d}$ of Eq. (\ref{maxDiamond}).
\end{customtheorem}
Since the bound is valid for any $M \ge 1$, its tightest value is achieved by maximizing over $M$. Remarkably, the difficult problem of adaptive multi-channel discrimination has been reduced to the discrimination of an ensemble of Choi matrices.
However, in general, the computation of the Helstrom limit $P_H\left(\{\rho_{\calE_n}^{\otimes uM}, p_n\}\right)$ may still be challenging and, for this reason, we may resort to further bounds.
In particular, by using bounds from Bures' fidelity $F(\rho,\sigma):=\tr \sqrt{\sqrt{\rho}\sigma\sqrt{\rho}}$, 
we can obtain a lower bound that is easier to evaluate~\cite{supp}
\be
P_u\ge P_{u,LB}^F=\sum_{k^\prime>k}p_{k^\prime}p_{k}F^{2uM}(\rho_{\calE_{k^\prime}},\rho_{\calE_k})-u\overline{\Delta}_M/2.
\label{bound1_F}
\ee
Below we consider symmetric cases where the bound of Theorem~\ref{theorem:LB} can be greatly simplified.


\textit{Ensembles with symmetries.}--~The general problem of adaptive multi-channel discrimination can be further simplified if the ensemble possesses certain symmetries. The first to consider is joint tele-covariance. A quantum channel $\calE$ is tele-covariant~\cite{holevo2002remarks,datta2006complementarity,zhuang2017additive,PLOB} when, for any teleportation unitary $U$ (e.g., Pauli operator) we may write ${\calE}(U \rho U^{\dagger}) = V {\calE}(\rho)V^{\dagger}$ for another generally-different unitary $V$. Then, an ensemble of channels $\{{\calE}_{k}\}$ is called jointly tele-covariant~\cite{pirandola2017ultimate,pirandola2018advances}, when we may write the condition of tele-covariance for all the elements of the ensemble \textit{and} the output unitary $V$ does not depend on the label $k$, i.e., it is universal for the ensemble.

For an ensemble of jointly tele-covariant channels, we may rewrite the previous universal simulation by choosing $\calT^M$ as the standard teleportation~\cite{bennett1992} applied to a single Choi matrix ($M=1$). Furthermore, this simulation is perfect, meaning that we have $\Delta_{\calE,1}=0$~\cite{PLOB}. As a result, Theorem~\ref{theorem:LB} reduces to $P_{u,LB}=P_H\left(\{\rho_{\calE_n}^{\otimes u}, p_n\}\right)$. Furthermore, this lower bound is achievable ($P_u = P_{u,LB}$) by probing the channels with $u$ copies of the maximally-entangled state $\zeta$, which also means that adaptive strategies are not needed for these channels. We have therefore automatically proved the following, which is a generalization of Ref.~\cite[Th.~3]{pirandola2017ultimate} from binary to multi-ary channel discrimination.
\begin{customcorollary}{1}\label{COROgen}
Consider arbitrary $m \ge 2$ jointly tele-covariant channels $\{\calE_n\}_{n=0}^{m-1}$ with prior probabilities $\{p_n\}_{n=0}^{m-1}$. The minimum error probability for their $u$-round adaptive discrimination equals the Helstrom limit computed over their Choi matrices
\be
P_{u} = P_H\left(\{\rho_{\calE_n}^{\otimes u}, p_n\}\right).\label{bound2tele}
\ee
This is achievable by a non-adaptive entanglement-based strategy
where $u$ copies of a maximally-entangled state $\zeta $ are sent through the
extended channel $\mathcal{E}_{n}\otimes \mathcal{I}$.
\end{customcorollary}
Examples of jointly tele-covariant channels are QECs and all Pauli channels, therefore including QDCs. By contrast, QADCs do not belong to this family.

We can perform another relevant simplification when the ensemble possesses GUS~\cite{cariolaro2010theory}, i.e., it has equal priors $p_n=1/m$ and the channels satisfy
$
\calE_n=S^{n} \calE_0 S^{\dagger n},
$
where the unitary $S^m$ equals identity. In this case, the Choi matrices $\rho_{\calE_n}^{\otimes uM}$ also have GUS with extended symmetry operators $S_{uM}=S^{\otimes uM}$. Then, the optimal POVM $\{\Pi_n\}_{n=0}^{m-1}$ for discriminating a GUS ensemble of states has the same type of symmetry, i.e., $\Pi_n=S_{uM}^{n} \Pi_0 S_{uM}^{\dagger n}$~\cite{cariolaro2010theory,dalla2015optimality}.
As a result, the lower bound in Theorem~\ref{theorem:LB} takes the form
\be
P_{u,LB}= 1-\frac{1}{2}u \Delta_{\calE_0,M}-\max_{\Pi_0}{\rm Tr}\left[\Pi_0 \rho_{\calE_0}^{\otimes uM}\right],
\label{LB_GUS}
\ee
where the maximization is constrained by POVM normalization condition.
Finally, if the channel ensemble has both the properties of GUS and joint tele-covariance, then we may write the ultimate achievable bound
\be
P_u= 1-\max_{\Pi_0}{\rm Tr}\left[\Pi_0 \rho_{\calE_0}^{\otimes u}\right].\label{POVMtoOPTIMIZE}
\ee
In the following, we consider CPF, which has the property of GUS as a natural symmetry.

{\em Channel position finding.---}
An important case where we have GUS is the problem of CPF (see Fig.~\ref{fig:schematic_general}(d) for a schematic). Consider an array of $m$ cells, each containing a channel acting on a $d_S-$dimensional subsystem $S_k$. The goal is to find the position $n$ of a target channel $\Phi^{(T)}$, knowing that all the other cells contain copies of a background channel $\Phi^{(B)}$. Formally, we consider equal-prior discrimination of $m$ multi-channels $\{\calE_n\}_{n=0}^{m-1}$, each expressed by
\be
\calE_n=\big(\otimes_{k\neq n} \Phi^{(B)}_{S_k}\big)\otimes \Phi^{(T)}_{S_n}.
\label{channel_GUS}
\ee
By taking $m$ maximally-entangled states at the input $\zeta^{\otimes m}$, we define the global Choi matrix of the multi-channel above, which has the following form
\be 
\rho_{\calE_n}
=\big(\otimes_{k\neq n}\left(\rho_{\Phi^{(B)}}\right)_{S_kI_k} \big)\otimes \left(\rho_{\Phi^{(T)}}\right)_{S_n I_n}.
\label{TMSV_out}
\ee

From the multi-channel $\calE_n$ we can derive an $M$-port PBT simulation $\calE_{n}^{M}$ by replacing each individual channel $\Phi^{(B/T)}$ with its $M$-port simulation. Correspondingly, the simulation error affecting the multi-channel is in terms of the errors associated to the simulation of the individual channels, i.e.,
$\Delta_{{\calE_{n},M}}=(m-1)\Delta_{{\Phi^{(B)},M}}+\Delta_{{\Phi^{(T)},M}} $~\cite{supp}. Because this expression is the same for any $n$, the average simulation error is simply
$
\overline{\Delta}_M=\sum_n p_n \Delta_{\calE_{n},M}=\Delta_{\calE_{0},M}
$.
Furthermore, from Eq.~(\ref{maxDiamond}) we have $\Delta_{\Phi^{(\ell)},M}\le \delta_{M,d_S}$, and we can write the simpler upper bound
$
\overline{\Delta}_M\le m\delta_{M,d_S}\sim md_S^2/M
$. The simulation error of the CPF problem can be used in previous equations. In particular, we can use it in Eq.~(\ref{bound1_F}) which here takes the form
\be
P_{u}\ge P_{u,LB}^F=\frac{m-1}{2m}F^{4uM}_{\Phi^{(B)},\Phi^{(T)}}-u\overline{\Delta}_M/2,
\label{bound_F_CPF}
\ee
where $F_{\Phi^{(B)},\Phi^{(T)}}$ 
is the fidelity between the Choi matrices of the target and background channels~\cite{footnote2}.

In order to show further applications of our theory, below we consider three families of channels, QECs, QDCs and QADCs. The first two are jointly tele-covariant, so that our Corollary~\ref{COROgen} and Eq.~(\ref{POVMtoOPTIMIZE}) can be applied.

\textit{Discrimination of erasure and depolarizing channels.---} Let us study the multi-ary discrimination of QECs and QDCs. 
Recall that the $d$-dimensional QEC with erasure probability $q$ can be written as $%
\mathcal{E}_{q}(\rho )=q\ketbra{e}{e}+(1-q)\rho $, where $\rho $ is the
input state and $\ketbra{e}{e}$ is a state living in an orthogonal space.
The $d$-dimensional QDC with depolarizing probability $q$ takes
instead the form $\mathcal{D}_{q}(\rho )=q\mathbb{I}_{d}+(1-q)\rho $, where $%
\mathbb{I}_{d}=d^{-1}I$ is the fully mixed state. These two types of
channels can be treated compactly by exploiting the formalism of the
orthogonal replacement channel. 
This is explained in detail in~\cite{supp}, where we also show that, for the special case of binary discrimination between QECs (or QDCs), we find exact analytical solutions for the ultimate error probability.

\begin{figure}
\centering
\includegraphics[width=0.47\textwidth]{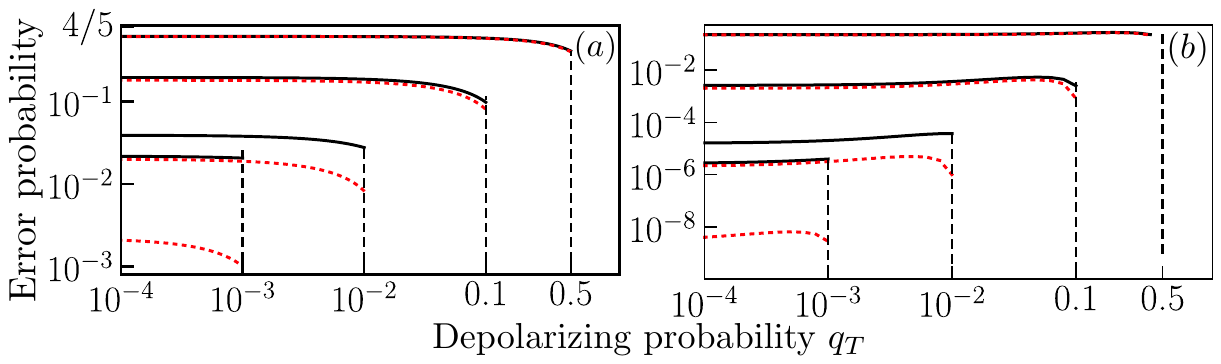}
\caption{Channel position finding with QDCs $\Phi
^{(B)}=\mathcal{D}_{q_{B}}$ and $\Phi ^{(T)}=\mathcal{D}_{q_{T}}$. We
consider $m=5$, $d=100$, and $q_{B}-q_{T}=0.5,0.9,0.99,0.999$ from top
to bottom. We compare the ultimate (entanglement-based) performance $P_{u}^{%
\text{QDC}}$ of Eq.~(\ref{QDCcpfMAIN}) (red curves) with the optimal
classical strategy based on un-entangled inputs (black curves). (a) $u=1$. (b) $u=3$. In all panels, the vertical dashed lines are
the maximum values that $q_T$ can take, because for those values we have $q_B=1$.}
\label{fig:pe_depo}
\end{figure}

Consider the multi-ary discrimination problem of CPF specified in Eq.~(\ref{channel_GUS}). Here
the background channel $\Phi ^{(B)}$ and the target channel $\Phi ^{(T)}$
are chosen to be QECs (or QDCs) with probabilities $q_{B}$ and $q_{T}$. For $%
m$ channels and $u$ uses, we define the function
\begin{eqnarray}
h_{m}^{u}\left( q_{B},q_{T}\right)  &:&=1-\frac{1}{m}\sum_{\bm x\in
\{0,1\}^{um}}\left[ q_{T}^{w^{\star }}(1-q_{T})^{u-w^{\star }}\times \right.
\notag  \label{PC_w_final_text} \\
&&\left. q_{B}^{\Vert \bm x\Vert -w^{\star }}(1-q_{B})^{(m-1)u-(\Vert \bm %
x\Vert -w^{\star })}\right] ,
\label{h_u_m}
\end{eqnarray}
where $w^{\star }=\max_{\ell }\Vert \bm x_{\ell }\Vert $ for $q_{T}\geq q_{B}
$, while $w^{\star }=\min_{\ell }\Vert \bm x_{\ell }\Vert $~\cite{footnote3} for $q_{T}<q_{B}$. Here $\bm x_\ell$ (with $0\le \ell \le m-1$) is the $(1+\ell u)$-th to $(\ell+1) u$-th components of the vector $\bm x$. Note that $h_{m}^{u}\left( q_{B},q_{T}\right)=h_{m}^{u}\left( 1-q_{B},1-q_{T}\right)$. Using this function, we compute $P_{u}$ in Eq.~(\ref{POVMtoOPTIMIZE}) and, when $u=1$, the summation can be simplified analytically~\cite{supp}.

For CPF\ with QECs $\Phi ^{(B)}=\mathcal{E}_{q_{B}}$ and $\Phi ^{(T)}=%
\mathcal{E}_{q_{T}}$, we find the ultimate error probability
\begin{equation}
P_{u}^{QEC}=h_{m}^{u}\left( q_{B},q_{T}\right) .
\end{equation}
In this case there is no
entanglement advantage, since we obtain the same performance by sending $u$\
copies of an optimal pure state $\phi ^{\otimes m}$ through $\mathcal{E}_{n}$ in a non-adaptive fashion.
For CPF with QDCs $\Phi ^{(B)}=\mathcal{D}_{q_{B}}$ and $\Phi
^{(T)}=\mathcal{D}_{q_{T}}$, we compute the ultimate error probability
\begin{equation}
P_{u}^{QDC}=h_{m}^{u}[ \left(1-{d^{-2}}\right) q_{T},\left( 1-{d^{-2}}%
\right) q_{B}].  \label{QDCcpfMAIN}
\end{equation}%
In this case, there is instead a clear advantage in using entanglement,
since the performance of an optimal pure state $\phi ^{\otimes m}$ is given by Eq.~(\ref{QDCcpfMAIN}) with the
replacement ${d^{-2}\rightarrow d^{-1}}$~\cite{supp}.  Fig.~\ref{fig:pe_depo} shows the gap between the entangled and non-entangled strategy which widens as the difference $|q_{B}-q_{T}|$ increases, and as the number of rounds $u$ increases. For one-shot discrimination ($u=1$) of a completely depolarizing channel $q_{T}=1$ among identity
channels ($q_{B}=0$), we may write $P_{1}^{QDC}={\left( m-1\right) }/{md^{2}}$~\cite{supp}.

\emph{Discrimination of amplitude damping channels.---} A QADC $\mathcal{A}%
_{q}$\ with damping probability $q$\ has Kraus decomposition $\mathcal{A}%
_{q}(\rho )=\sum_{i=0,1}K_{i}\rho K_{i}^{\dagger }$, with operators $K_{0}:=%
\ketbra{0}{0}+\sqrt{1-q}\ketbra{1}{1}$ and $K_{1}:=\sqrt{q}\ketbra{0}{1}$. It
is not tele-covariant and its PBT simulation has non-zero error $\Delta _{%
\mathcal{A}_{q},M}=\xi _{M}[\left( 1-q\right) /2+\sqrt{1-q}],$ where $\xi
_{M}$ is the constant given in Ref.~\cite[Eq.~(11)]{pirandola2019fundamental}. 
While the binary discrimination
between two QADCs has been treated in the literature~\cite{pirandola2019fundamental} (see~\cite{supp} for further results on receiver designs and pretty-good measurement (PGM)~\cite{PGM1,PGM2,PGM3}), little is known in the setting of multi-ary discrimination.

\begin{figure}
\includegraphics[width=0.47\textwidth]{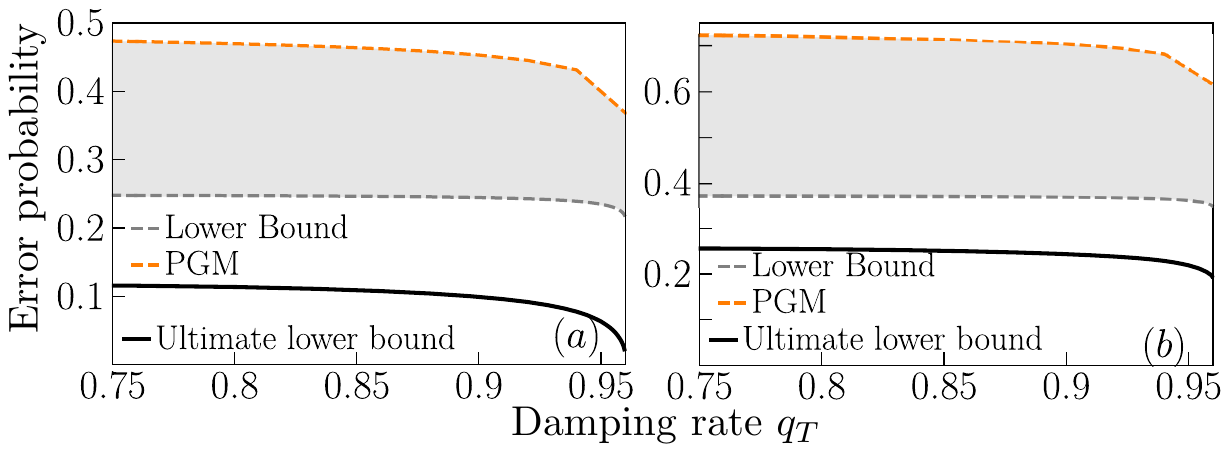}
\caption{Channel position finding with QADCs $\mathcal{A}_{q_{B}}$ and $%
\mathcal{A}_{q_{T}}$ for $q_{B}=q_{T}+0.04$. The solid black curve is the
ultimate lower bound $P_{u,LB}^{F\star}$ optimized from Eq.~(\ref{bound_CPF_damping}). The non-adaptive performance lies between the fidelity
lower bound (gray dashed curve) and the PGM upper bound (orange dashed
curve) as given by Eq.~(\ref{LBeUB}). (a) We consider $m=2$ and $u=4$%
. (b) We consider $m=4$ and $u=2$.}
\label{QADCfigures}
\end{figure}

Consider the multi-ary discrimination problem of CPF specified in Eq.~(\ref{channel_GUS}), with background $\Phi ^{(B)}=\mathcal{A}_{q_{B}}$
and target $\Phi ^{(T)}=\mathcal{A}_{q_{T}}$. We compute the lower
bound in Eq.~(\ref{bound_F_CPF}) here taking the form
\begin{equation}
P_{u}\geq P_{u,LB}^{F}=\frac{m-1}{2m}F^{4uM}-u\overline{\Delta }_{M}/2,
\label{bound_CPF_damping}
\end{equation}%
where $F:=\left[ 1+\sqrt{(1-q_{B})(1-q_{T})}+\sqrt{q_{B}q_{T}}\right] /2$
and $\overline{\Delta }_{M}=(m-1)\Delta _{\mathcal{A}_{q_{B}}{,M}}+\Delta _{%
\mathcal{A}_{q_{T}}{,M}}$. By optimizing over $M$, we derive its tightest form $P_{u,LB}^{F\star }=\max_{M}P_{u,LB}^{F}$. As a comparison, we
consider a non-adaptive scheme, where $u$ copies of the maximally entangled
state $\zeta ^{\otimes m}$ probe $\mathcal{E}_{n}$. Correspondingly, the Helstrom limit computed on the ensemble of output Choi matrices $\{\rho _{\mathcal{E}_{n}}^{\otimes u}\} $ is bounded as~\cite{supp}
\begin{equation}
\frac{m-1}{2m}F^{4u}\leq P_{H}\left( \{\rho _{\mathcal{E}_{n}}^{\otimes
u},1/m\}\right) \leq P_{E}^{\mathrm{PGM}},  \label{LBeUB}
\end{equation}%
where $P_{E}^{\mathrm{PGM}}$ is the performance achievable via a PGM at the output~\cite{PGM1,PGM2,PGM3}. Fig.~\ref{QADCfigures} shows a gap between the ultimate lower bound $P_{u,LB}^{F\star}$ and the non-adaptive performance. Further investigation
is needed to establish if this gap is effectively due to adaptiveness.

{\em Conclusions.---} In this work, we established the ultimate limits for the minimum error probability affecting the (generally-adaptive) statistical discrimination of an arbitrary $m\ge2$ number of finite-dimensional quantum channels. We find remarkable simplifications in the presence of symmetries, with our bound becoming exactly achievable when the channel ensemble is jointly tele-covariant. Our theory allows us to find the ultimate performances achievable in the fundamental m-ary discrimination problem of CPF, considering various types of channels. In particular, for CPF with depolarizing channels, we show that the use of entanglement greatly outperforms the performance of any classical strategy.

Note that CPF can be translated into various applications, including readout of memories, radar scanning and absorbance spectroscopy.
For instance, CPF may model the readout process from a digital memory where information is encoded in the position of a target cell within a block. In the frequency domain, this is equivalent to finding the absorbance line within a spectrum.
A possible future direction is developing our theory in the setting of unambiguous hypothesis testing, suitably extending Refs.~\cite{Janos2010,Janos2002,USD1,Janos2005,USD2} to $m$-ary channel discrimination.

\begin{acknowledgements}
Q.Z. acknowledges funding from Army Research Office under Grant No. W911NF-19-1-0418, Office of Naval Research under Grant No. N00014-19-1-2189, Defense Advanced Research Projects Agency (DARPA) under Young Faculty Award (YFA) Grant No. N660012014029, and University of Arizona. S.P. acknowledges funding from the European Union's Horizon 2020 Research and Innovation Action under grant agreement No. 862644 (Quantum readout techniques and technologies, QUARTET). 
\end{acknowledgements}

%

\newpage 

\
\newpage 

\begin{widetext}
{
\centering 
\bf \large
Supplementary Material: Ultimate limits for multiple quantum channel discrimination
}
\\


\tableofcontents

\

In this Supplementary Material, we present detailed proofs for various results presented in the main paper, and we also provide further theoretical results and analyses. In Sec.~\ref{sup_unitaries}, we apply the ultimate limit to unitaries with geometric uniform symmetry (GUS). In Sec.~\ref{proof:lemma:helstrom_continuity}, we prove Lemma~\ref{lemma:helstrom_continuity} of the main paper. In Sec.~\ref{proof:theorem:LB}, we prove Theorem~\ref{theorem:LB} of the main paper. In Sec.~\ref{App:bounds}, we provide various bounds for the Helstrom limit. In Sec.~\ref{App:simulation_error}, we bound the simulation error for the problem of channel position finding (CPF). In Sec.~\ref{ORCappendix}, we employ the orthogonal replacement channel (ORC) to perform analysis for quantum erasure channels (QECs) and quantum depolarizing channels (QDPs). This analysis includes new results for the binary discrimination of these channels, before treating the corresponding performances in the $m$-ary problem of CPF. In Sec.~\ref{QADCappendix}, we present new results for the binary discrimination of qubit amplitude damping channels (QADCs), including the performance achievable by using a pretty-good measurement (PGM) and a nulling receiver.

To facilitate the readers, we list all acronyms in table~\ref{table_acr} below.
\begin{table}[H]
\begin{center}
\begin{tabular}{ |c|c| } 
 \hline
quantum channel discrimination &QCD\\
 \hline
geometric uniform symmetry &GUS\\
\hline
channel position finding &CPF\\
\hline
quantum erasure channel&QEC\\
\hline
quantum depolarizing channel &QDC\\
 \hline
qubit amplitude damping channel& QADC\\
 \hline
positive-valued operator measure & POVM\\
 \hline
port-based teleportation & PBT\\
 \hline
pretty-good measurement &PGM\\
 \hline
orthogonal replacement channel &ORC \\
 \hline
\end{tabular}
 \caption{A list of acronyms in this paper.\label{table_acr}}
\end{center}
\end{table}

\section{GUS unitaries}
\label{sup_unitaries}

In the case of GUS unitaries $\{U_n\}_{n=0}^{m-1}$, the Choi matrices
\begin{equation}
\rho_{\mathcal{E}_n}=\left(U_n\otimes I\right) \zeta
\left(U_n^\dagger\otimes I\right)
\end{equation}
are pure states. The lower bound $P_{u}\ge P_{u,LB}=P_H-u\overline{\Delta}_M/2$ in Theorem~\ref{theorem:LB} of the main text consists of a Helstrom bound part and a simulation error part. The Helstrom bound in Eq.~(\ref{HelstromFIRST}) of the main text can be solved analytically
\begin{equation}
P_H=\frac{m-1}{m^2}\left[\sqrt{1+(m-1)\eta}-\sqrt{1-\eta}\right]^2,
\label{P_H_PPM}
\end{equation}
where $\eta=\tr \left[\zeta \left(U_1^\dagger S U_1 \otimes I\right)\right]^{u}$. Note that this lower bound $P_{u,LB}$ does not conflict with the fact that any
two unitaries can be perfectly distinguished when $u$ is large but finite~\cite{acin2001statistical}. 
This is due to the fact that an ensemble of unitaries does not have, in general, the property of joint tele-covariance and, therefore the lower bound always has non-zero
simulation error $u\overline{\Delta}_M>0$. As $u$ increases, $P_{u,LB}$ can become negative in the general case. Comparing the threshold of such a positive-to-negative transition with the minimum number of rounds for zero error discrimination in Ref.~\cite{acin2001statistical} will be an interesting further task. 

\section{Proof of Lemma~\ref{lemma:helstrom_continuity} of the main paper}

\label{proof:lemma:helstrom_continuity}

\begin{proof}
First the Helstrom limit can be written as the optimization of the POVM
elements $\{\Pi_n^\prime\}_{n=0}^{m-1}$, each corresponding to a hypothesis $n$,
\begin{align}
P_H\left(\{\rho_n^\prime, p_n\}\right)&:=1-\max_{\sum_n
\Pi_n^\prime=I}\sum_n p_n \mathrm{Tr}\left(\rho_n^\prime \Pi_n^\prime\right)
\\
&=1-\sum_n p_n \mathrm{Tr}\left(\rho_n^\prime \Pi_n^{\prime\star}\right)
\label{lemma1_proof_eq1}
\\
&=1-\sum_n p_n \mathrm{Tr}\left(\rho_n \Pi_n^{\prime\star}\right)-\sum_n p_n
\mathrm{Tr}\left[\left(\rho_n^\prime-\rho_n\right) \Pi_n^{\prime\star}\right]
\\
&\ge P_H\left(\{\rho_n, p_n\}\right)-\frac{1}{2}\sum_n p_n
\|\rho_n^\prime-\rho_n\| 
\label{lemma1_proof_ineq1}
\\
&\ge P_H\left(\{\rho_n, p_n\}\right)-\frac{1}{2}\sum p_n \delta_n.
\end{align}
In Eq.~(\ref{lemma1_proof_eq1}), $\{\Pi_n^{\prime\star}\}_{n=0}^{m=1}$ is the optimum POVM achieving $P_H\left(\{\rho_n^\prime, p_n\}\right)$. In Ineq.~(\ref{lemma1_proof_ineq1}), we used
the fact that $1-\sum_n p_n \mathrm{Tr}\left(\rho_n
\Pi_n^{\prime\star}\right)$ is the error probability for hypothesis testing
on the ensemble $\{\rho_n\}$ with prior probability distribution $\{p_n\}$, using the POVM $\{\Pi_n^{\prime\star}\}_{n=0}^{m-1}$; therefore the error probability cannot beat the
Helstrom limit; we have also used one-norm's variational form
\begin{equation}
\|A\|=2\sup_{0\le P\le I} \mathrm{Tr}\left[P A\right],
\end{equation}
so that $\mathrm{Tr}\left[\left(\rho_n^\prime-\rho_n\right)
\Pi_n^{\prime\star}\right]\le \|\rho_n^\prime-\rho_n\|/2$.
\end{proof}

\section{Proof of Theorem~\ref{theorem:LB} of the main paper}

\label{proof:theorem:LB}

\begin{proof}
For any $u$-round adaptive protocol, from Ineq.~(\ref{stretchingEQ}) in the main paper, we have the output states $\rho_{%
\mathcal{E}_n,u}$
\begin{equation}
\|\rho_{\mathcal{E}_n,u}-\Lambda\left(\rho_{\mathcal{E}_n}^{\otimes
uM}\right)\| \le u \Delta_{\mathcal{E}_n,M}.
\end{equation}
The Helstrom limit $P_H\left(\{\rho_{\mathcal{E}_n,u}, p_n\}\right)$ of
these output states $\{\rho_{\mathcal{E}_n,u}\}$ with prior distribution $\{p_n\}$ gives
the performance of the protocol. From Lemma~\ref{lemma:helstrom_continuity} of the main paper, we have
\begin{align}
&P_H\left(\{\rho_{\mathcal{E}_n,u}, p_n\}\right)  \notag \\
&\ge P_H\left(\{\Lambda\left(\rho_{\mathcal{E}_n}^{\otimes
uM}\right), p_n\}\right)-\frac{1}{2}\sum_n p_n u \Delta_{\mathcal{E}_n,M}
\\
&\ge P_H\left(\{\rho_{\mathcal{E}_n}^{\otimes uM}, p_n\}\right)-\frac{1}{2}\sum_n p_n u \Delta_{\mathcal{E}_n,M} 
\label{theorem2_proof_step}
\\
&=P_H\left(\{\rho_{\mathcal{E}_n}^{\otimes uM}, p_n\}\right)-\frac{1}{2}
u \overline{\Delta}_M.
\end{align}
where $\overline{\Delta}_M=\sum_n p_n \Delta_{\mathcal{E}_n,M}$. In Ineq.~(\ref{theorem2_proof_step}), we have used data-processing inequality in hypothesis testing. 
Note that we can simply replace $\Delta_{\mathcal{E}_n,M}$ by $\delta_{M,d}$, thus $\overline{\Delta}_M$ in the final bound can be replaced
by $\delta_{M,d}$ due to its independence from the channel.
\end{proof}

\section{General bounds}

\label{App:bounds}

Here we discuss various general bounds for the Helstrom limit $P_H\left(\{\rho_n, p_n\}\right)$, which is known to be difficult to compute. An upper bound can be obtained from the pretty good measurement (PGM)~\cite{PGM1,PGM2,PGM3}
described by the POVM
\begin{equation}
\Pi^{\mathrm{PGM}}_n=\Sigma^{-1/2}p_n\rho_n \Sigma^{-1/2},~~~0\le n \le m-1,
\end{equation}
where $\Sigma=\sum_{n=0}^{m-1}p_n \rho_n$. Clearly, $\sum_{n=0}^{m-1} \Pi^{%
\mathrm{PGM}}_n=I$ and each element is positive. The error probability is
therefore
\begin{equation}
P_E^{\mathrm{PGM}} =1-\sum_{n=0}^{m-1} p_n \tr\left(\Pi^{\mathrm{PGM}%
}_n\rho_n\right)\ge P_H\left(\{\rho_n, p_n\}\right).
\end{equation}
Ref.~\cite{Barnum} gives a further upper bound
\begin{equation}
P_H\leq P_{H,UB}:= 2 \sum_{k^\prime>k}\sqrt{p_{n^\prime}p_{n}}%
F(\rho_{n^\prime},\rho_{n}),  \label{Barnum}
\end{equation}
where $F$ is the Bures' fidelity
\begin{equation}
F(\rho,\sigma):=\Vert\sqrt{\rho}\sqrt{\sigma}\Vert_{1} =\tr \sqrt{\sqrt{\rho}%
\sigma\sqrt{\rho}}.
\end{equation}
A fidelity-based lower bound is instead given by~\cite{montanaro2008lower},
\begin{equation}
P_H\geq P_{H,LB}:=
\sum_{n^\prime>n}p_{n^\prime}p_{n}F^{2}(\rho_{n^\prime},\rho_{n}).
\label{montanaro2008lower}
\end{equation}

Assume equi-probable hypotheses, so that $p_{n}=m^{-1}$ for any $n$, and the
symmetry $F(\rho_n,\rho_{n^\prime})=F$, $\forall n\neq n^\prime$. We then have
the simplified bounds
\begin{align}
P_{H,UB}:= (m-1)F,  \label{UB2} \\
P_{H,LB}:= \frac{m-1}{2m}F^2.  \label{LB2}
\end{align}

Since the CPF problem has GUS, if we consider a GUS product
input $\otimes_{k=1}^m\phi_{S_k}$, the output state becomes
\begin{equation}
\rho_n=\big(\otimes_{k\neq n} \sigma^{(B)}_{S_k}\big) \otimes
\sigma^{(T)}_{S_n},  \label{state_GUS}
\end{equation}
where $\sigma^{(T/B)}:=\Phi^{(T/B)}(\phi)$. It is clear that this ensemble
of output states also has GUS, i.e., $\rho_n=S^n \rho_0 S^{\dagger n}, $ and
it is analogous to the states in a PPM~\cite%
{yuen1975optimum,eldar2004optimal,cariolaro2010theory}.

Therefore for CPF problem, we have
\begin{equation}
F(\rho_n,\rho_{n^\prime\neq n})=F^2\big(\sigma^{(T)},\sigma^{(B)}\big).
\end{equation}
In the main paper, where each channel $\Phi^{(B/T)}$
is extended to $\left(\Phi^{(B/T)}\otimes \mathcal{I}\right)^{\otimes uM}$,
the state $\sigma^{(B/T)}$ is replaced by the Choi matrix $\rho_{\Phi^{(B/T)}}^{%
\otimes uM}$. This leads to
\begin{equation}
F(\rho_n,\rho_{n^\prime\neq n})=F^2\big(\rho_{\Phi^{(B)}}^{\otimes
uM},\rho_{\Phi^{(T)}}^{\otimes uM}\big)=F^{2uM}\big(\rho_{\Phi^{(B)}},\rho_{%
\Phi^{(T)}}\big).
\end{equation}

\section{Bound on the simulation error}
\label{App:simulation_error}

\label{App:CPF_error} In the CPF problem, each multi-channel is described by
\begin{equation}
\mathcal{E}_n=\big(\otimes_{k\neq n} \Phi^{(B)}_{S_k}\big)\otimes
\Phi^{(T)}_{S_n},~~~0\le n \le m-1,
\end{equation}
where $\Phi^{(B/T)}_{S_k}$ is the background/target channel acting on $d_S-$%
dimensional subsystem $S_k$. The error between $\mathcal{E}_n$ and its $M$-port PBT simulation $\mathcal{E}_n^M$ can
be bounded by the error between each $\Phi^{(B/T)}$ and its PBT simulation $\Phi^{(B/T),M}$,
\begin{align}
&\Delta_{\mathcal{E}_n,M}:= \|\mathcal{E}_n-\mathcal{E}_n^M\|_\diamond
\notag \\
&=\|\big(\otimes_{k\neq n} \Phi^{(B)}_{S_k}\big)\otimes \Phi^{(T)}_{S_n}-%
\big(\otimes_{k\neq n} \Phi^{(B),M}_{S_k}\big)\otimes
\Phi^{(T),M}_{S_n}\|_\diamond  \notag \\
&\le
(m-1)\|\Phi^{(B)}-\Phi^{(B),M}\|_\diamond+\|\Phi^{(T)}-\Phi^{(T),M}\|_%
\diamond  \notag \\
&=(m-1)\Delta_{\Phi^{(B)},M}+\Delta_{\Phi^{(T)},M},
\end{align}
where we used $\|\Phi_1\otimes
\Phi_2-\Phi_1^\prime\otimes\Phi_2^\prime\|_\diamond \le
\|\Phi_2-\Phi_2^\prime\|_\diamond+\|\Phi_1-\Phi_1^\prime\|_\diamond$
repeatedly. Because of the GUS property of the multi-channel ensemble
considered in CPF, we have that the expression above also holds for the
average simulation error, i.e.,
\begin{equation}
\overline{\Delta}_M=\sum_n p_n \Delta_{\mathcal{E}_{n},M}=(m-1)\Delta_{{%
\Phi^{(B)},M}}+\Delta_{{\Phi^{(T)},M}}.  \label{Delta_CPFappendix}
\end{equation}
Clearly, we can also set $\overline{\Delta}_M=\Delta_{\mathcal{E}_{0},M}$.


\section{Ultimate limits for quantum erasure channels and quantum depolarizing channels\label{ORCappendix}}
In this section, we apply Corollary~\ref{COROgen} of the main paper to calculate the ultimate lower bound for QECs and QDPs. In particular, we develop the orthogonal replacement channel (ORC) as a tool for our analysis. Overall, the results in this section is summarized in the following lemmas and propositions, which we will prove in the following subsections.


\begin{customlemma}{S1}\label{lemma_ORC_binary}
Consider the binary discrimination between two ORCs $\mathcal{R}_{q_0,\rho ^{\perp }}$ and $\mathcal{R}_{q_1,\rho ^{\perp }}$, where $\mathcal{R}_{q,\rho ^{\perp }}(\rho )=q\rho^{\perp } +(1-q)\rho $ with $\rho $ as the input and $\rho ^{\perp }$ as some state in an
orthogonal space. The Helstrom limit between outputs from arbitrary pure state input $\phi^{\otimes u}$ in $u$ channel uses is given by
\begin{equation}
P_{H}(u,\phi
)=f_{u}\left( q_{0},q_{1}\right) :=\frac{1}{2}-\frac{1}{4}%
\sum_{k=0}^{u}C_{u}^{k}|q_{0}^{k}(1-q_{0})^{u-k}-q_{1}^{k}(1-q_{1})^{u-k}|,
\end{equation}
where $C_{u}^{k}$ is the binomial coefficient. In particular, for the one-shot binary discrimination ($u=1$), we can write 
\be
f_{1}\left( q_{0},q_{1}\right) =\left( 1-|q_{0}-q_{1}|\right) /2.
\ee
\end{customlemma}

\begin{customproposition}{S1.1}\label{corollary_binary_QEC}
Given two QECs, $\mathcal{E}_{q_{0}}$ and $\mathcal{E}_{q_{1}}$, where $\mathcal{E}_{q}(\rho )=q\ketbra{e}{e}+(1-q)\rho$, the minimum error probability for their $u$-round adaptive discrimination equals
\begin{equation}
P_u^{QEC}=f_{u}\left( q_{0},q_{1}\right),
\end{equation}
and neither adaptiveness nor entanglement is necessary to achieve this.
\end{customproposition}

\begin{customproposition}{S1.2}\label{corollary_binary_QDC}
Given two QDCs, $\mathcal{D}_{q_{0}}$ and $\mathcal{D}_{q_{1}}$, where $\mathcal{D}_{q}(\rho )=q\mathbb{I}_{d}+(1-q)\rho$, the minimum error probability for their $u$-round adaptive discrimination equals
\begin{equation}
P_u^{QDC}=f_{u}[\left( 1-{d^{-2}}\right) q_{0},\left( 1-{%
d^{-2}}\right) q_{1}].
\end{equation}%
To achieve this optimal performance adaptiveness is not needed but a maximally entangled input $\zeta^{\otimes u}$ is necessary.
\end{customproposition}

\begin{customlemma}{S2}\label{lemma_ORC_CPF}
Consider the multi-ary discrimination problem of CPF specified in Eq.~(\ref{channel_GUS}) in the main paper. Here
the background channel $\Phi ^{(B)}$ and the target channel $\Phi ^{(T)}$
are chosen to be $\mathcal{R}_{q_B,\rho ^{\perp }}$ and $\mathcal{R}_{q_T,\rho ^{\perp }}$. For $%
m$ channels and $u$ uses with arbitrary pure input $\phi^{\otimes um}$, the Helstrom limit between the output states is given by
\begin{eqnarray}
P_{H}(u,m,\phi )=h_{m}^{u}\left( q_{B},q_{T}\right)  &:&=1-\frac{1}{m}\sum_{\bm x\in
\{0,1\}^{um}}\left[ q_{T}^{w^{\star }}(1-q_{T})^{u-w^{\star }} q_{B}^{\Vert \bm x\Vert -w^{\star }}(1-q_{B})^{(m-1)u-(\Vert \bm x\Vert -w^{\star })}\right],
\end{eqnarray}
where $w^{\star }=\max_{\ell }\Vert \bm x_{\ell }\Vert $ for $q_{T}\geq q_{B}
$, while $w^{\star }=\min_{\ell }\Vert \bm x_{\ell }\Vert $ for $q_{T}<q_{B}$. Here $\bm x_\ell$ (with $0\le \ell \le m-1$) is the $(1+\ell u)$-th to $(\ell+1) u$-th components of the vector $\bm x$. When $u=1$, the summation can be analytically solved to give
\begin{align}
& h_{m}^{1}\left( q_{B},q_{T}\right) := 1-\frac{1}{m}\left[
q_{T}q_{B}^{m-1}+(1-q_{T})(1-q_{B})^{m-1} +\left( 1-(1-q_{B})^{m}-q_{B}^{m}\right) \max \left( \frac{q_{T}}{%
q_{B}},\frac{1-q_{T}}{1-q_{B}}\right) \right]. 
\end{align}

\end{customlemma}

\begin{customproposition}{S2.1} \label{corollary_CPF_QEC}
For CPF\ with QECs $\Phi ^{(B)}=\mathcal{E}_{q_{B}}$ and $\Phi ^{(T)}=%
\mathcal{E}_{q_{T}}$, the minimum error probability for their $u$-round adaptive discrimination equals
\begin{equation}
P_{u}^{QEC}=h_{m}^{u}\left( q_{B},q_{T}\right),
\end{equation}
and neither adaptiveness nor entanglement is necessary to achieve this.
\end{customproposition}

\begin{customproposition}{S2.2} \label{corollary_CPF_QDC}
For CPF with QDCs $\Phi ^{(B)}=\mathcal{D}_{q_{B}}$ and $\Phi
^{(T)}=\mathcal{D}_{q_{T}}$, the minimum error probability for their $u$-round adaptive discrimination equals
\begin{equation}
P_{u}^{QDC}=h_{m}^{u}[\left( 1-{d^{-2}}\right) q_{T},\left( 1-{d^{-2}}%
\right) q_{B}]. 
\end{equation}%
To achieve this optimal performance adaptiveness is not needed 
but a maximally entangled input $\zeta^{\otimes u}$ is necessary. For one-shot discrimination, the formula greatly simplifies. In particular, for $q_{B}=0$ and $q_{T}=1$, we have $P_{1}^{QDC}={\left(
m-1\right) }/{md^{2}}$, while, for $q_{B}=1$ and $q_{T}=0$, we have $P_{1}^{QDC}\simeq {\left( m-1\right) }/{2d^{2}}$.
\end{customproposition}

\begin{customremark}{S1}
In the error probability functions of lemmas~\ref{lemma_ORC_binary} and \ref{lemma_ORC_CPF}, one has the symmetry of $q\leftrightarrow 1-q$. Namely,
\begin{align} 
f_{u}\left( q_{0},q_{1}\right)&=f_{u}\left( 1-q_{0},1-q_{1}\right)
\\
h_{m}^{u}\left( q_{B},q_{T}\right)&=h_{m}^{u}\left( 1-q_{B},1-q_{T}\right).
\end{align} 
This also agrees with the intuition of the symmetry between $q\rho^{\perp } +(1-q)\rho$ and $q\rho +(1-q)\rho^{\perp }$ in terms of hypothesis testing.
\end{customremark}

\subsection{Preliminary definitions}

We define the ORC with replacement
probability $q$ and state $\rho ^{\perp }$ as
\begin{equation}
\mathcal{R}_{q,\rho ^{\perp }}(\rho )=q\rho^{\perp } +(1-q)\rho,
\end{equation}%
where\ $\rho $ is the input and $\rho ^{\perp }$ is some fixed state in an
orthogonal space. The $d$-dimensional QEC with erasure probability $q$ can
be written as
\begin{equation}
\mathcal{E}_{q}(\rho )=q\ketbra{e}{e}+(1-q)\rho=\mathcal{R}_{q,%
\ketbra{e}{e}}\left( \rho \right) ,  \label{QEC}
\end{equation}%
for any state $\rho $, where the orthogonal state $\rho^\perp=\ketbra{e}$. Denote $\mathbb{I}_{d}:=I/d$ as the fully mixed state.
Then, the $d$-dimensional QDC with depolarizing probability $q$ can be
written as 
\be 
\mathcal{D}_{q}(\rho )=q\mathbb{I}_{d}+(1-q)\rho.
\ee 
For a fixed
pure input state $\phi $, we may write the output state \be 
\mathcal{D}%
_{q}(\phi )=\mathcal{R}_{q_{d},\mathbb{I}_{(d-1)}}(\phi )
\ee 
with $%
q_{d}=\left( 1-d^{-1}\right) q$ and $\rho ^{\perp }=\mathbb{I}_{(d-1)}$
is the fully mixed state acting on the $(d-1)$-dimensional Hilbert space
orthogonal to the input $\phi $. Note that this equality only holds in terms
of the output (it does not mean that the channels are equal).

Similarly, consider the output of the extended ORC $\mathcal{R}_{q,\rho
^{\perp }}\otimes \mathcal{I}(\rho _{SI})=q\rho ^{\perp
}\otimes \rho _{I}+(1-q)\rho _{SI}=\mathcal{R}_{q,\rho ^{\perp }\otimes \rho _{I}}(\rho
_{SI}),$ where the orthogonal state $\rho ^{\perp }\otimes \rho _{I}$ lives
in a larger Hilbert space and also depends on the input. In particular, for
a maximally entangled state $\zeta $ at the input of a QEC\ and a QDC, we
may respectively write the output states as follow%
\begin{eqnarray}
\mathcal{E}_{q}\otimes \mathcal{I}(\zeta ) &=&\mathcal{R}_{q,\ketbra{e}{e}%
\otimes \mathbb{I}_{d}}(\zeta ),  \label{QEC_extend} \\
\mathcal{D}_{q}\otimes \mathcal{I}(\zeta ) &=&\mathcal{R}_{q_{d^{2}},\mathbb{%
I}_{(d^{2}-1)}}(\zeta ).  \label{QDC_extend}
\end{eqnarray}

\subsection{Binary discrimination}
In this section, we will prove Lemma~\ref{lemma_ORC_binary} and Propositions~\ref{corollary_binary_QEC} and \ref{corollary_binary_QDC}.

Consider the binary discrimination between $\mathcal{R}_{q_{0}}$ and $%
\mathcal{R}_{q_{1}}$ with equal priors (where, in the notation, we have
omitted the orthogonal state for simplicity). All calculations for the
binary case reduce to the calculation of the Helstrom limit 
\be 
P_{H}(u,\phi
):=\frac{1}{2}\left( 1-\frac{1}{2}\Vert \mathcal{R}_{q_{0}}(\phi )^{\otimes u}-\mathcal{R}%
_{q_{1}}(\phi )^{\otimes u}\Vert \right) 
\ee 
for a pure state input $\phi $. Let us use the expansion
\begin{equation}
\mathcal{R}_{q}(\phi )^{\otimes u}=\sum_{\bm x}q^{\Vert \bm x\Vert
}(1-q)^{u-\Vert \bm x\Vert }\rho _{\bm x},
\end{equation}%
where the state $\rho _{\bm x}=\otimes _{n}\left( \sigma _{n}\right) _{S_{n}}
$ is indexed by a vector $\bm x\in \{0,1\}^{u}$, and we have $\sigma
_{n}=\rho ^{\perp }$ for $x_{n}=1$, and $\sigma _{n}=\phi $ when $x_{n}=0$.
For example, for $u=3$, possible states could be%
\begin{eqnarray}
\rho _{(1,0,0)} &=&\rho _{S_{0}}^{\perp }\otimes \phi _{S_{1}}\otimes \phi
_{S_{2}}, \\
\rho _{(1,0,1)} &=&\rho _{S_{0}}^{\perp }\otimes \phi _{S_{1}}\otimes \rho
_{S_{2}}^{\perp }.
\end{eqnarray}%
We note that the possible states $\rho _{\bm x}$ are in orthogonal supports,
for any pure input state $\phi $. In other words, we may write
\begin{equation}
\mathrm{Tr}\left( \rho _{\bm x}\rho _{\bm x^{\prime }}\right) =\delta _{\bm %
x=\bm x^{\prime }}\mathrm{Tr}(\rho ^{\perp 2})^{\Vert \bm x\Vert }.
\end{equation}%
This observation directly allows us to solve the binary case. In fact, we
may write
\begin{align}
& \Vert \mathcal{R}_{q_{0}}(\phi )^{\otimes u}-\mathcal{R}_{q_{1}}(\phi
)^{\otimes u}\Vert   \notag \\
& =\Vert \sum_{\bm x}\left( q_{0}^{\Vert \bm x\Vert }(1-q_{0})^{u-\Vert \bm %
x\Vert }-q_{1}^{\Vert \bm x\Vert }(1-q_{1})^{u-\Vert \bm x\Vert }\right)
\rho _{\bm x}\Vert  \\
&
=\sum_{k=0}^{u}C_{u}^{k}|q_{0}^{k}(1-q_{0})^{u-k}-q_{1}^{k}(1-q_{1})^{u-k}|:=g_{u}(q_{0},q_{1}),
\end{align}%
where $C_{u}^{k}$ is the binomial coefficient. Therefore, we can write%
\begin{equation}
P_{H}(u,\phi )=f_{u}\left( q_{0},q_{1}\right) :=\frac{1}{2}\left[ 1-\frac{1}{%
2}g_{u}(q_{0},q_{1})\right]=\frac{1}{2}-\frac{1}{4}%
\sum_{k=0}^{u}C_{u}^{k}|q_{0}^{k}(1-q_{0})^{u-k}-q_{1}^{k}(1-q_{1})^{u-k}|,
\label{binary_qrc}
\end{equation}
which is our claim in Lemma~\ref{lemma_ORC_binary}. Note that, for $u=1$, it takes the simple form 
\be 
f_{1}\left( q_{0},q_{1}\right) =\left( 1-|q_{0}-q_{1}|\right) /2.
\ee

Given two QECs $\mathcal{E}_{q_{0}}$ and $\mathcal{E}_{q_{1}}$, the ORC form
in Eq.~(\ref{QEC}) leads to the result
\begin{equation}
P_{H}^{QEC}(u,\phi )=f_{u}\left( q_{0},q_{1}\right) .
\end{equation}
Note that an entangled state $\zeta $ leads to the same result, i.e., we have
\begin{equation}
P_{H}^{QEC}(u,\zeta)=P_{H}^{QEC}(u,\phi).
\end{equation}
This is due to the specific extended ORC form in
Eq.~(\ref{QEC_extend}), which has the same probability of the ORC in Eq.~(\ref{QEC}). The same error probability is
achieved by sending $u$\ copies of an optimal single-system pure state $\phi
$ through $\mathcal{E}_{q}$, or equivalently by sending $u$ copies of a maximally
entangled state $\zeta $\ through the extended channel $\mathcal{E}%
_{q}\otimes \mathcal{I}$. In other words, the optimal performance is achievable by 
strategies without entanglement. Finally, recall from Corollary~3 of the main text that, for these channels, adaptiveness is not needed and the ultimate performance  $P_u^{QEC}$ is equal to $P_{H}^{QEC}(u,\zeta)$. As a result, we have proven the claims 
of our Proposition~\ref{corollary_binary_QEC}.

\begin{figure*}
\centering
\includegraphics[width=1\textwidth]{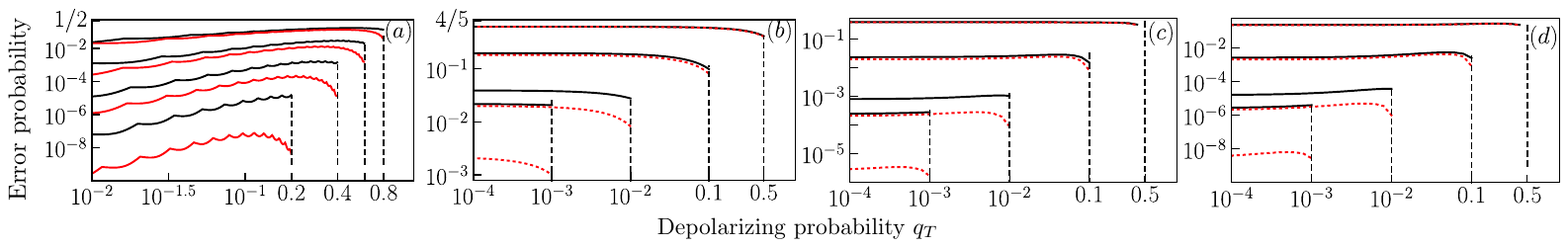}
\caption{Quantum channel discrimination with depolarizing channels. (a)
Binary discrimination between $\mathcal{D}_{q_{0}}$ and $\mathcal{D}_{q_{1}}$
for $u=30$ rounds, $d=6$ dimension, and $q_{0}-q_{1}=0.2,0.4,0.6,0.8$ from
top to bottom. We compare the ultimate (entanglement-based) performance (Eq.~\ref{QDCperf}, red curves)
with the optimal classical strategy based on un-entangled inputs (Eq.~\ref{QDCclassical}, black
curves).\ (b)~Channel position finding with depolarizing channels $\Phi
^{(B)}=\mathcal{D}_{q_{B}}$ and $\Phi ^{(T)}=\mathcal{D}_{q_{T}}$. We
consider $u=1$, $m=5$, $d=100$, and $q_{B}-q_{T}=0.5,0.9,0.99,0.999$ from top
to bottom. We compare the ultimate (entanglement-based) performance $P_{u}^{%
\text{QDC}}$ in (Eq.~\ref{CPF_depo_E}, red curves) with the optimal
classical strategy based on un-entangled inputs (Eq.~\ref{CPF_depo_C}, black curves). (c) Same as
(b) but $u=2$. (d) Same as (b) but $u=3$. In all panels, the vertical dashed lines are
the maximum values that $q_T$ can take, because for those values we have $q_B=1$.}
\label{fig:pe_depo}
\end{figure*}

Given two QDCs $\mathcal{D}_{q_{0}}$ and $\mathcal{D}_{q_{1}}$, the ORC form
in Eq.~(\ref{QDC_extend}) gives
\begin{equation}
P_{H}^{QDC}(u,\zeta )=f_{u}[\left( 1-{d^{-2}}\right) q_{0},\left( 1-{%
d^{-2}}\right) q_{1}].
\label{QDCperf}
\end{equation}%
For comparison, without entanglement, we need to consider the ORC form $%
\mathcal{D}_{q}(\phi )=\mathcal{R}_{q_{d},\mathbb{I}_{(d-1)}}(\phi )$ which
leads to the error probability
\begin{equation}
P_{H}^{QDC}(u,\phi )=f_{u}[\left( 1-{d}^{-1}\right) q_{0},\left( 1-{d}%
^{-1}\right) q_{1}].
\label{QDCclassical}
\end{equation}
We see that the different dimensions in the probability of the ORC ($q_{d^{2}}
$ versus $q_{d}$) leads to a performance difference in the error probabilities [$%
P_{H}^{QDC}(u,\zeta )$ versus $P_{H}^{QDC}(u,\phi )$]. Again recall from Corollary~3 of the main text that, for these channels too, adaptiveness is not needed, and the ultimate performance $P_u^{QDC}$ is equal to $P_{H}^{QDC}(u,\zeta)$. As a result, we have proven the claims of our Proposition~\ref{corollary_binary_QDC}.

We plot the results in Fig.~\ref{fig:pe_depo}(a), where we see a clear gap between the ultimate
entanglement-based performance in Eq.~(\ref{QDCperf}) and the classical
strategy without entanglement in Eq.~(\ref{QDCclassical}). This gap widens as the difference $q_{0}-q_{1}$
increases. In fact, one can show for any $u$, the advantage is
largest when $|q_{0}-q_{1}|=1$, where the error probability with and without
entanglement scales as $1/2d^{2u}$ and $1/2d^{u}$, respectively.

\subsection{Channel position finding}
In this section, we prove Lemma~\ref{lemma_ORC_CPF} and Propositions~\ref{corollary_CPF_QEC} and \ref{corollary_CPF_QDC}.

Now we consider the CPF problem with target channel $\mathcal{R}_{q_{T}}$
and background channel $\mathcal{R}_{q_{B}}$. In order words, we consider a
problem of $m$-ary channel discrimination where, with the same prior
probability $p_{n}=m^{-1}$, we have a generic multi-channel $\mathcal{E}_{n}=%
\big(\otimes _{k\neq n}\Phi _{S_{k}}^{(B)}\big)\otimes \Phi _{S_{n}}^{(T)}$
with $\Phi ^{(B)}=\mathcal{R}_{q_{B}}$ and $\Phi ^{(T)}=\mathcal{R}_{q_{T}}$%
. For an arbitrary pure input state $\phi $, let us consider the possible
equiprobable output states $\rho _{n}^{\otimes u}=[\mathcal{E}_{n}\left(
\phi ^{\otimes m}\right) ]^{\otimes u}$ after $u$ uses of the multi-channel $%
\mathcal{E}_{n}$. Our goal is to compute the Helstrom limit $P_{H}(\{\rho
_{n},p_{n}\})$ following Eq.~(\ref{HelstromFIRST}) of the main text.

First of all we find that $\rho _{n}=\sum_{\bm x}g(\bm x,n)\rho _{\bm x}$,
with $\bm x\in \{0,1\}^{um}$ and coefficients
\begin{align}
& g(\bm x,n)=   q_{T}^{\Vert \bm x_{n}\Vert }(1-q_{T})^{u-\Vert \bm x_{n}\Vert
}\prod_{k\neq n}q_{B}^{\Vert \bm x_{k}\Vert }(1-q_{B})^{u-\Vert \bm %
x_{k}\Vert },
\end{align}%
where each $\bm x_{k}\in \{0,1\}^{u}$ represents the state in the subsystem $%
k$, for $0\leq k\leq m-1$.

By making use of the GUS $\mathcal{E}_{n}=S^{n}\mathcal{E}_{0}S^{\dagger n}$%
, we can simplify the Helstrom limit to the form
\begin{equation}
P_{H}(u,m,\phi )=1-\max_{\Pi _{0}}\mathrm{Tr}\left[ \Pi _{0}\rho
_{0}^{\otimes u}\right] ,  \label{PHorc}
\end{equation}%
where $\Pi _{0}$\ is a POVM operator and constrained by normalization. Note that Eq.~(\ref{PHorc}) becomes $%
P_{u}$ of Eq.~(\ref{POVMtoOPTIMIZE}) of the main text, when we extend the
channel $\mathcal{E}_{n}\left( \phi ^{\otimes m}\right) \rightarrow (%
\mathcal{E}_{n}\otimes \mathcal{I})(\zeta ^{\otimes m})$, so that $\rho
_{0}^{\otimes u}$ becomes the Choi matrix $\rho_{\mathcal{E}_{0}}^{\otimes
u} $.

As we show in Sec.~\ref{App:CPF_ORC}, we may compute Eq.~(\ref{PHorc}). For
any pure input state $\phi $, we obtain%
\begin{equation}
P_{H}(u,m,\phi )=h_{m}^{u}\left( q_{B},q_{T}\right) :=1-\frac{1}{m}\sum_{\bm %
x\in \{0,1\}^{um}}g^{\star }(\bm x,n),  \label{PC_w_final2}
\end{equation}%
where $g^{\star }(\bm x,n)=\max_{k\in \lbrack 0,m-1]}g(S^{k}\bm x,n)$. One
can further solve the maximization and obtain
\begin{align}
g^{\star }(\bm x,n) &=q_{T}^{w^{\star }}(1-q_{T})^{u-w^{\star }}  q_{B}^{\Vert \bm x\Vert
-w^{\star }}(1-q_{B})^{(m-1)u-(\Vert \bm x\Vert -w^{\star })},
\end{align}
where $w^{\star }=\max_{\ell }\Vert \bm x_{\ell }\Vert $ for $q_{T}\geq
q_{B} $, while $w^{\star }=\min_{\ell }\Vert \bm x_{\ell }\Vert $ for $q_{T}<q_{B}$. This is the main claim of our Lemma~\ref{lemma_ORC_CPF}. Note that, for the specific case of one-shot discrimination $u=1$, we can obtain an analytical solution to the
summation in Eq.~(\ref{PC_w_final2}), finding
\begin{align}
& h_{m}^{1}\left( q_{B},q_{T}\right) := 1-\frac{1}{m}\left[
q_{T}q_{B}^{m-1}+(1-q_{T})(1-q_{B})^{m-1} +\left( 1-(1-q_{B})^{m}-q_{B}^{m}\right) \max \left( \frac{q_{T}}{%
q_{B}},\frac{1-q_{T}}{1-q_{B}}\right) \right] ,  \label{PH_ORC_CPF}
\end{align}%
for $q_{B}<1$. For $q_{B}=1$, we instead have $h_{m}^{1}\left(
q_{B},q_{T}\right) =(m-1)q_{T}/m$ (see Sec.~\ref{App:CPF_ORC}, including the summation in Sec.~\ref{CPF_u1_details}, for more
technical details).

Let us now specify the ORCs to QECs and QDCs. For CPF with QECs $\mathcal{E}%
_{q_{B}}$ and $\mathcal{E}_{q_{T}}$, there is no entanglement advantage
(similar to the binary discrimination case). In fact, we compute
\begin{equation}
P_{H}^{QEC}(u,m,\phi )=h_{m}^{u}\left( q_{B},q_{T}\right) ,
\end{equation}
for any pure input $\phi $, and we find
\begin{equation}
P_{H}^{QEC}(u,m,\zeta )=P_{H}^{QEC}(u,m,\phi ),
\end{equation}
when we extend the channel to a maximally entangled input $\zeta $. Combining this result with Corollary~3 of the main text, we prove the main claim for the 
ultimate error probability $P_{u}^{QEC}$ which is stated in our Proposition~\ref{corollary_CPF_QEC} (and also reported in the discussions of our main text).

For CPF with QDCs $\mathcal{D}_{q_{B}}$ and $\mathcal{D}_{q_{T}}$, we
compute the performance without entanglement
\begin{equation}
P_{H}^{QDC}(u,m,\phi )=h_{m}^{u}[\left( 1-{d}^{-1}\right) q_{T},\left( 1-%
{d}^{-1}\right) q_{B}],
\label{CPF_depo_C}
\end{equation}%
and the ultimate limit achieved by entangled strategy
\begin{eqnarray}
P_{H}^{QDC}(u,m,\zeta )  = h_{m}^{u}[\left( 1-{d^{-2}}\right) q_{T},\left( 1-{d^{-2}}\right)
q_{B}],
\label{CPF_depo_E}
\end{eqnarray}%
with clear advantage in the presence of entanglement (similar to the binary
discrimination case). 
Combining this result with Corollary~3 of the main text, we prove the main claim for the ultimate error probability $P_{u}^{QDC}$ which is stated in our Proposition~\ref{corollary_CPF_QDC} (and also reported in the discussions of our main text). Then, in Fig.~\ref{fig:pe_depo}(b)-(d) we provide the comparison between the ultimate limit achieved by the entangled strategy in Eq.~(\ref{CPF_depo_E}) versus the classical strategy performance in Eq.~(\ref{CPF_depo_C}).

\subsection{More details on CPF with orthogonal replacement channels\label{App:CPF_ORC}}

The output state of an ORC can be written as
\begin{equation}
\mathcal{R}_{q}(\phi )^{\otimes u}=\sum_{\bm x}q^{\Vert \bm x\Vert
}(1-q)^{u-\Vert \bm x\Vert }\rho _{\bm x},
\end{equation}%
where the state $\rho _{\bm x}=\otimes _{n}\left( \sigma _{n}\right)
_{S_{n}} $ is indexed by a vector $\bm x\in \{0,1\}^{u}$, $\sigma _{n}=\rho
^{\perp }$ when $x_{n}=1$ and $\sigma _{n}=\phi $ when $x_{n}=0$. Consider
CPF with background channel $\Phi ^{(B)}=\mathcal{R}_{q_{B}}$ and target
channel $\Phi ^{(T)}=\mathcal{R}_{q_{T}}$. The states at the output of $u$
uses of the generic multi-channel $\mathcal{E}_{n}$ are given by
\begin{equation}
\rho _{n}^{\otimes u}=[\mathcal{E}_{n}\left( \phi ^{\otimes m}\right)
]^{\otimes u}=\sum_{\bm x}g(\bm x,n)\rho _{\bm x},
\end{equation}%
where $\bm x\in \{0,1\}^{um}$ and
\begin{align}
& g(\bm x,n)=   q_{T}^{\Vert \bm x_{n}\Vert }(1-q_{T})^{u-\Vert \bm x_{n}\Vert
}\prod_{k\neq n}q_{B}^{\Vert \bm x_{k}\Vert }(1-q_{B})^{u-\Vert \bm %
x_{k}\Vert },  \label{gfun_app}
\end{align}%
where each $\bm x_{k}\in \{0,1\}^{u}$, $0\leq k\leq m-1$.

For the CPF case, we need more analyses by choosing a set of complete and
orthonormal bases. Moreover, if we consider the bases $\ket{\phi^\perp_k}$
such that $\rho ^{\perp }=\sum_{k=0}^{d^{\perp }-1}\lambda _{k}%
\ketbra{\phi^\perp_k}{\phi^\perp_k}$ is diagonal, then we have $\rho _{\bm %
x} $ diagonal in bases formed by products of $\ket{\phi^\perp_k}$ and $%
\ket{\phi}$. We denote each bases projector as $A_{\bm x}$, which satisfy
the normalization
\begin{equation}
\mathrm{Tr}\left( A_{\bm x}A_{\bm x^{\prime }}^{\prime }\right) =\delta _{%
\bm x=\bm x^{\prime }}\delta _{A=A^{\prime }}.
\end{equation}%
The notation $\mathcal{A}_{\bm x}$ is the set of projectors that act on the
Hilbert space that $\rho _{\bm x}$ lives in. As an example, $m=3$ and $u=1$
case, $\rho _{(1,0,0)}$ is diagonal in bases
\begin{equation}
\mathcal{A}_{(1,0,0)}=\{\ketbra{\phi^\perp_k}{\phi^\perp_k}_{S_{0}}\otimes %
\ketbra{\phi}{\phi}_{S_{1}}\otimes \ketbra{\phi}{\phi}_{S_{2}},0\leq k\leq
d^{\perp }-1\}.
\end{equation}%
Note that in this way $S^{n}A_{\bm x}S^{n\dagger }=A_{S^{n}\bm x}$.

Then all states $\rho _{n}$ are diagonal, thus we only need to consider projective measurements in the corresponding bases.
Similar to the analysis in Eq.~(\ref{LB_GUS}), one can further consider the
GUS projective POVM $\Pi _{n}=S^{n}\Pi _{0}S^{\dagger n}$, with
\begin{equation}
\Pi _{0}=\sum_{\bm x}\sum_{A_{\bm x}\in \mathcal{A}_{\bm x}}\lambda _{A_{\bm %
x}}A_{\bm x},  \label{GUS_expansion}
\end{equation}%
where positivity requires $\lambda _{A_{\bm x}}\geq 0$. We define the
ensemble $\mathbb{S}_{\bm x}=\{S^{-n}\bm x\}_{n=0}^{m-1}$ for later use.
Note that we define $\mathbb{S}_{\bm x}$ such that it always have $m$
elements, although there are elements that repeat the others. We can also
define the set version $\mathbb{\tilde{S}}_{\bm x}$, where members don't
repeat. For instance, $\mathbb{S}_{(1,0,1,0)}=\{(1,0,1,0),(0,1,0,1),(1,0,1,0),(0,1,0,1)\}$
has 4 members, but if we consider the ensemble as a set, then there are only
2 members, i.e., $\mathbb{\tilde{S}}_{(1,0,1,0)}=\{(1,0,1,0),(0,1,0,1)\}$ .

We directly have
\begin{equation}
\mathrm{Tr}\left(A_{\bm x} \rho_{\bm x^\prime}\right)=\delta_{\bm x=\bm %
x^\prime} \mathrm{Tr}\left(A_{\bm x} \rho_{\bm x}\right).
\end{equation}
Moreover, $\sum_{A_{\bm x}\in \mathcal{A}_{\bm x}} \mathrm{Tr}\left(A_{\bm %
x} \rho_{\bm x}\right)=1$.

Completeness requires
\begin{align}
& \sum_{n=0}^{m-1}S^{n}\Pi _{0}S^{n\dagger }=\sum_{n=0}^{m-1}\sum_{\bm %
x}\sum_{A_{\bm x}\in \mathcal{A}_{\bm x}}\lambda _{A_{\bm x}}S^{n}A_{\bm %
x}S^{n\dagger }  =\sum_{\bm x}\sum_{A_{\bm x}\in \mathcal{A}_{\bm x}}(\sum_{n=0}^{m-1}%
\lambda _{A_{S^{-n}\bm x}})A_{\bm x}=I.
\end{align}%
Therefore, we have the normalization $\sum_{n=0}^{m-1}\lambda _{A_{S^{-n}\bm x}}=1$
as $A_{\bm x}$'s are projectors. Equivalently we may write
\begin{equation}
\sum_{\bm y\in \mathbb{S}_{\bm x}}\lambda _{A_{\bm y}}=1,\forall \bm x.
\label{lambda_normalization}
\end{equation}

The probability of making a correct decision is
\begin{align}
& P_{C}=\mathrm{Tr}\left( \Pi _{0}\rho _{0}^{\otimes u}\right)  \\
& =\sum_{\bm x}\sum_{A_{\bm x}\in \mathcal{A}_{\bm x}}g(\bm x,0)\lambda _{A_{%
\bm x}}\mathrm{Tr}\left( \rho _{\bm x}A_{\bm x}\right) 
\label{PC_first}
\\
& =\sum_{\mathbb{\tilde{S}}_{\bm x}}\sum_{\bm y\in \mathbb{\tilde{S}}_{\bm %
x}}\sum_{A_{\bm y}\in \mathcal{A}_{\bm y}}\mathrm{Tr}\left( \rho _{\bm y}A_{%
\bm y}\right) g(\bm y,0)\lambda _{A_{\bm y}}.
\label{PC_second}
\end{align}%
Note that in the first summation, we only sum over different sets $\mathbb{%
\tilde{S}}_{\bm x}$. Because for $\bm y\in \mathbb{\tilde{S}}_{\bm x}$, $%
\mathrm{Tr}\left( A_{\bm y}\rho _{\bm y}\right) =\mathrm{Tr}\left( A_{\bm %
x}\rho _{\bm x}\right) $ does not depend on $\bm y$, the maximum is achieved
when $\lambda _{A_{\bm x^{\star }}}=1$, where $\bm x^{\star }=\arg \max_{\bm %
y\in \mathbb{\tilde{S}}_{\bm x}}g(\bm y,0)$. Thus
\begin{equation}
P_{C}^{\star }=\sum_{\mathbb{\tilde{S}}_{\bm x}}\frac{|\mathbb{\tilde{S}}_{%
\bm x}|}{|\mathbb{S}_{\bm x}|}g(\bm x^{\star },0)=\sum_{\mathbb{\tilde{S}}_{%
\bm x}}\frac{|\mathbb{\tilde{S}}_{\bm x}|}{m}g(\bm x^{\star },0).
\label{PC_app}
\end{equation}%
The pre-factor comes from the fact that the normalization in Eq.~(\ref{lambda_normalization}) is for $\mathbb{S}_{\bm x}$ instead of $\mathbb{%
\tilde{S}}_{\bm x}$. Because the sets $\mathbb{\tilde{S}}_{\bm x}$ are
non-overlapping and covers all possible $\bm x\in \{0,1\}^{um}$, we can
simply write
\begin{equation}
P_{C}^{\star }=\frac{1}{m}\sum_{\bm x}g_{w}(w_{\mathrm{min}},w_{\mathrm{max}%
},W),  \label{PC_app_w_final}
\end{equation}%
where $g_{w}(w_{\mathrm{min}},w_{\mathrm{max}},W)=g(\bm x^{\star },0)$ is
explained as follows. Denote $w_{\ell }=\Vert \bm x_{\ell }\Vert $, and $w_{%
\mathrm{max}}=\max_{\ell }\Vert \bm x_{\ell }\Vert $, $w_{\mathrm{min}%
}=\min_{\ell }\Vert \bm x_{\ell }\Vert $ and $W=\sum_{\ell }w_{\ell }$.
Recall the form of $g$ function in Eq.~(\ref{gfun_app}). Then, we have
\begin{align}
& g(\bm x^{\star },0)=\max_{\bm y\in \mathbb{S}_{\bm x}}g(\bm y,0)  =g_{w}(w_{\mathrm{min}},w_{\mathrm{max}},W):=  \left\{
\begin{array}{l}
q_{T}^{w_{\mathrm{max}}}(1-q_{T})^{u-w_{\mathrm{max}}}q_{B}^{W-w_{\mathrm{max}}}(1-q_{B})^{(m-1)u-(W-w_{%
\mathrm{max}})} \\
\ \ \ \ \ \ \ \ \ \ \ \ \ \ \ \ \ \ \mbox{if $q_T\ge q_B$}; \\
q_{T}^{w_{\mathrm{min}}}(1-q_{T})^{u-w_{\mathrm{min}}}q_{B}^{W-w_{\mathrm{min}}}(1-q_{B})^{(m-1)u-(W-w_{%
\mathrm{min}})} \\
\ \ \ \ \ \ \ \ \ \ \ \ \ \ \ \ \ \ \mbox{if $q_T<q_B$}.%
\end{array}%
\right.   \label{gfun_simp_app}
\end{align}%
We see the only dependence is on the maximum, minimum and total weights.

\subsubsection{A simple example}

Here we give a simple example that is helpful to understand the notation. Consider CPF with qubit QECs for $u=1$ and $m=2$. Suppose that there is a pure input $\ket{\phi}$ and the replacement pure state is $\ket{e}$. Let's denote the two states as $\ket{0},%
\ket{1}$ for simplicity. So now the four projectors are $A_{00}=%
\ketbra{00}{00},A_{10}=\ketbra{10}{10},A_{01}=\ketbra{01}{01},A_{11}=%
\ketbra{11}{11}$ (each set $\mathcal{A}_{\bm x}$ only has one member). We
can decompose the POVM as in Eq.~(\ref{GUS_expansion})
\begin{equation}
\Pi _{0}=\lambda _{00}A_{00}+\lambda _{01}A_{01}+\lambda _{10}A_{10}+\lambda
_{11}A_{11}.
\end{equation}%
The normalization condition gives
\begin{align}
& \sum_{n=0}^{1}S^{n}\Pi _{0}S^{n\dagger }=\lambda _{00}A_{00}+\lambda
_{01}A_{01}+\lambda _{10}A_{10}+\lambda _{11}A_{11}  \notag \\
& +\lambda _{00}A_{00}+\lambda _{01}A_{10}+\lambda _{10}A_{01}+\lambda
_{11}A_{11}=I.
\end{align}%
This leads to the normalization as in Eq.~(\ref{lambda_normalization})
\begin{equation}
\lambda _{00}+\lambda _{00}=\lambda _{11}+\lambda _{11}=\lambda
_{01}+\lambda _{10}=1.
\end{equation}%
The ensembles are $\mathbb{S}_{00}=\{00,00\},\mathbb{S}_{11}=\{11,11\},\mathbb{S}%
_{10}=\mathbb{S}_{01}=\{10,01\}$ and their set versions are $\mathbb{\tilde{S}}%
_{00}=\{00\},\mathbb{\tilde{S}}_{11}=\{11\},\mathbb{\tilde{S}}_{10}=%
\mathbb{\tilde{S}}_{01}=\{10,01\}$. There are only three different ensembles.

We first evaluate the coefficients from Eq.~\ref{gfun_app}
\begin{align}
& g(00,0)=q_{T}^{0}(1-q_{T})^{1}q_{B}^{0}(1-q_{B})^{1}=(1-q_{T})(1-q_{B}), \\
& g(01,0)=q_{T}^{0}(1-q_{T})^{1}q_{B}^{1}(1-q_{B})^{0}=(1-q_{T})q_{B}, \\
& g(10,0)=q_{T}^{1}(1-q_{T})^{0}q_{B}^{0}(1-q_{B})^{0}=q_{T}(1-q_{B}), \\
& g(11,0)=q_{T}^{1}(1-q_{T})^{0}q_{B}^{1}(1-q_{B})^{0}=q_{T}q_{B}.
\end{align}%
We consider the case with $q_{T}\geq q_{B}$. From Eq.~(\ref{PC_first}), the correct probability is therefore
\begin{eqnarray}
P_{C} =g(00,0)\lambda _{00}+g(01,0)\lambda _{01}  +g(10,0)\lambda _{10}+g(11,0)\lambda _{11},
\end{eqnarray}%
where we used $\mathrm{Tr}\left( \rho _{00}A_{00}\right) =\mathrm{Tr}\left(
\rho _{01}A_{01}\right) =\mathrm{Tr}\left( \rho _{10}A_{10}\right) =\mathrm{Tr}\left( \rho _{11}A_{11}\right) =1$. In the second way of writing in Eq.~(\ref{PC_second}), we sum
over ensembles $\mathbb{\tilde{S}}_{00},\mathbb{\tilde{S}}_{10},\mathbb{\tilde{S}}_{01}$ and can obtain the same result. Now we consider the maximum
correct probability, from Eq.~(\ref{PC_app}) or Eq.~(\ref{PC_app_w_final})
\begin{eqnarray}
P_{C}^{\star } =\frac{1}{2}q_{T}q_{B}+\frac{1}{2}(1-q_{T})(1-q_{B})  +\max \{q_{T}(1-q_{B}),(1-q_{T})q_{B}\},
\end{eqnarray}%
which agrees with the direct intuition.

\subsubsection{Analytical results for CPF with $u=1$}
\label{CPF_u1_details}
For $u=1$, except for $\bm x=\bm0,\bm1$, we always have $w_{\mathrm{min}}=0,w_{\mathrm{max}}=1$. We consider the various cases.

\begin{enumerate}
\item $W=0$, we have $w_{\mathrm{max}}=w_{\mathrm{min}}=0$, thus $%
g_w(0,0,0)=(1-q_T) (1-q_B)^{m-1}$.

\item $W=m$, we have $w_{\mathrm{max}}=w_{\mathrm{min}}=1$; the contribution
is $g_w(1,1,m)=q_Tq_B^{m-1}$.

\item $0<W<m$, we have two possibilities
\begin{equation}
g_w(0,1,W)=\left\{
\begin{array}{ll}
q_T q_B^{W-1}(1-q_B)^{m-W}  & \mbox{if $q_T \geq q_B$}, \\
(1-q_T)q_B^{W}(1-q_B)^{m-1-W} & \mbox{if $q_T < q_B$}.%
\end{array}
\right.
\end{equation}
\end{enumerate}

Overall the maximum correctness probability from Eq.~(\ref{PC_app_w_final}) is
\begin{align}
& P_{C}^{\star }=q_{T}q_{B}^{m-1}/m+(1-q_{T})(1-q_{B})^{m-1}/m+   \left\{
\begin{array}{ll}
\sum_{k=1}^{m-1}\frac{C_{m}^{k}}{m}q_T q_B^{k-1}(1-q_B)^{m-k} &
\mbox{if $q_T
\geq q_B$}; \\
\sum_{k=1}^{m-1}\frac{C_{m}^{k}}{m}(1-q_T)q_B^{k}(1-q_B)^{m-1-k} &
\mbox{if
$q_T < q_B$}.%
\end{array}%
\right.
\end{align}%
In the above formula, we have used the fact that when the weight is fixed to be $W$, there are only $C_{m}^{W}$ possible vectors $\bm x$.
The summations can be performed to give Eq.~(\ref{PH_ORC_CPF}).

We compare the analytical and numerical results for $u=1$ in Fig.~\ref%
{fig:numerics_analytical}(a), and we see exact agreement; we also used
two numerical methods, where they agree in Fig.~\ref{fig:numerics_analytical}(b) for the $u>1$ case.

\begin{figure}[tbp]
\centering
\subfigure{\includegraphics[width=0.25\textwidth]{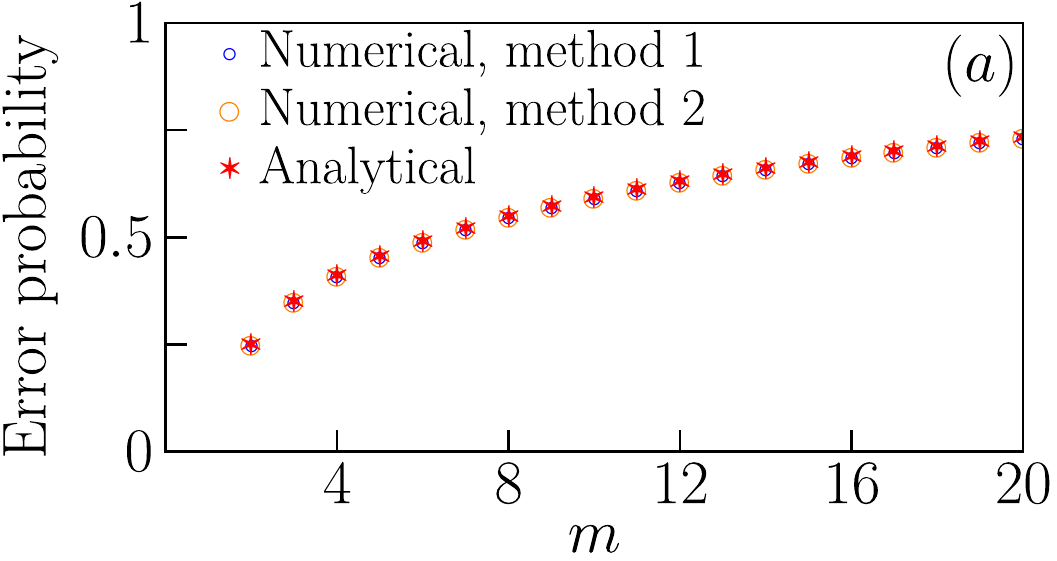}
\label{fig:numerical_analytical_u1_m}
} \subfigure{\includegraphics[width=0.25\textwidth]{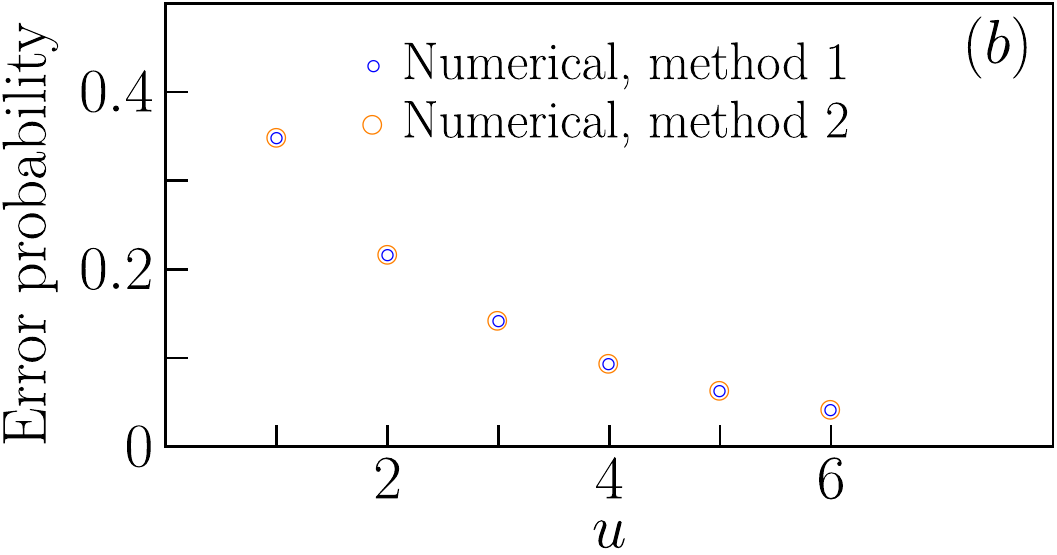}
\label{fig:numerical_m3_u}
}
\caption{Error probability of CPF between ORCs with $q_T=0.4,q_B=0.9$. Numerical method 1 is
based on Eq.~(\ref{PC_app}) and numerical method 2 is based on Eq.~(\ref{PC_app_w_final}). (a) Comparison between numerical approaches
and analytical results in Eq.~(\ref{PH_ORC_CPF}) of the main paper
for $u=1$. (b) Numerical results for $m=3$. }
\label{fig:numerics_analytical}
\end{figure}


\section{Binary discrimination of amplitude damping channels\label{QADCappendix}}

The binary discrimination with equal priors between $\mathcal{A}_{q_{0}}$
and $\mathcal{A}_{q_{1}}$ has been treated in Ref.~\cite{pirandola2019fundamental}. As summarized in Fig.~\ref{dampingFIG}, here we perform additional analyses via a nulling receiver design, the PGM bound and the numerical evaluation of the Helstrom limit. 

Consider a non-adaptive protocol, where $u$
copies of the maximally-entangled state $\zeta $ probe the unknown channel $\mathcal{A}_{q}$. This strategy provides $u$ copies of the Choi matrix $\rho
_{\mathcal{A}_{q}}$ at the output, so that we need to discriminate between
the two equiprobable Choi matrices $\rho _{\mathcal{A}_{q_{0}}}^{\otimes u}$
and $\rho _{\mathcal{A}_{q_{1}}}^{\otimes u}$. For the corresponding
non-adaptive Helstrom limit, we can apply the Fuchs-van de Graaf relations~\cite{FuchsGraaf} and write the following upper and lower bounds
\begin{equation}
\frac{1-\sqrt{1-F^{2u}}}{2}\leq P_{H}\left( \{\rho _{\mathcal{A}%
_{q_{0}}}^{\otimes u},\rho _{\mathcal{A}_{q_{1}}}^{\otimes u}\}\right) \leq
\frac{F^{u}}{2},  \label{P_H_binary_bounds}
\end{equation}%
where $F=\left[ 1+\sqrt{(1-q_{0})(1-q_{1})}+\sqrt{q_{0}q_{1}}\right] /2$ is
the fidelity between the two Choi matrices.

\begin{figure}[tbp]
\centering
\includegraphics[width=0.35\textwidth]{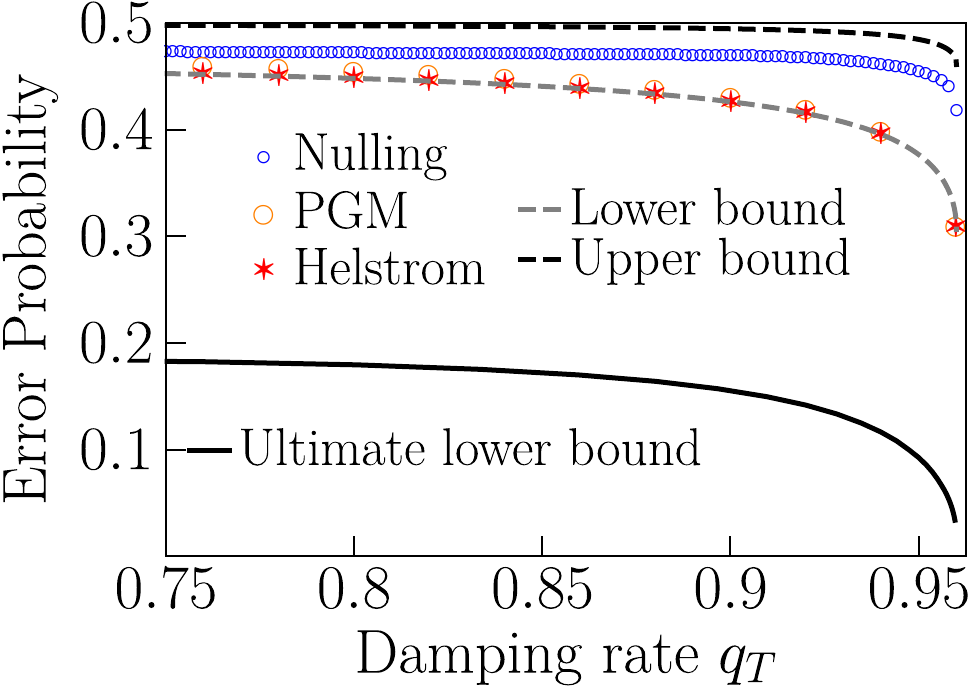} 
\caption{Error probability for the binary discrimination of QADCs $\mathcal{A%
}_{q_{0}}$ and $\mathcal{A}_{q_{1}}$ with $q_{0}=q_{1}+0.04$. We consider $%
u=8$ rounds. The black solid curve is the ultimate lower bound $%
P_{u,LB}^{F\star }$. Then, we compare the nulling strategy (blue circles),
the PGM strategy (orange circles), the non-adaptive Helstrom limit (red
stars), and the lower (gray dashed line) and upper bound (black dashed line)
in Eq.~(\ref{P_H_binary_bounds}). }
\label{dampingFIG}
\end{figure}

By specifying Eq.~(\ref{bound2}) of the main text to the binary case and using the lower
bound in Eq.~(\ref{P_H_binary_bounds}), we may write the following lower
bound for adaptive discrimination
\begin{equation}
P_{u}\geq P_{u,LB}^{F}=\frac{1-u\overline{\Delta }_{M}-\sqrt{1-F^{2uM}}}{2},
\label{P_u_F_binary}
\end{equation}%
where $\overline{\Delta }_{M}=\Delta _{\mathcal{A}_{q_{0}},M}+\Delta _{\mathcal{A}_{q_{1}},M}$. By maximizing over the number of ports $M$ one
obtains the ultimate lower bound $P_{u,LB}^{F^{\star }}=\max_{M}P_{u,LB}^{F}$, as given in Ref.~\cite{pirandola2019fundamental}. It is interesting to
compare this performance with that of two non-adaptive strategies: the
nulling strategy (explained below) and the PGM
strategy~\cite{PGM1,PGM2,PGM3}(see Sec.~\ref{App:bounds} for details on PGM). The results for $u=9$ rounds and damping
rate $q_{0}=q_{1}+0.04$ are shown in Fig.~\ref{dampingFIG}, where we see
that the PGM, the non-adaptive Helstrom limit and its lower bound in Eq.~(\ref{P_H_binary_bounds}) lie on top of each other. The nulling strategy,
while being based on local operations, is better than the upper bound in
Eq.~(\ref{P_H_binary_bounds}).

\subsection{Details of the nulling strategy\label{Appendix_nulling}}

The Choi matrix in the computational basis takes the form
\begin{equation}
\rho _{\mathcal{A}_{q}}=\left(
\begin{array}{cccc}
\frac{1}{2} & 0 & 0 & \frac{\sqrt{1-q}}{2} \\
0 & \frac{q}{2} & 0 & 0 \\
0 & 0 & 0 & 0 \\
\frac{\sqrt{1-q}}{2} & 0 & 0 & \frac{1-q}{2}
\end{array}%
\right) .
\end{equation}%
The nulling strategy originates from the observation that one can find
unitary
\begin{equation}
U_{q}=\left(
\begin{array}{cccc}
-\sqrt{\frac{1-q}{2-q}} & 0 & 0 & \frac{1}{\sqrt{2-q}} \\
0 & 0 & 1 & 0 \\
\frac{1}{\sqrt{2-q}} & 0 & 0 & \sqrt{\frac{1-q}{2-q}} \\
0 & 1 & 0 & 0%
\end{array}%
\right),
\end{equation}%
such that
\begin{equation}
U_{q}\rho _{\mathcal{A}_{q}}U_{q}^{\dagger }=\mathrm{Diag}\left[ 0,0,1-\frac{q}{2},-\frac{q}{2}\right]
\end{equation}%
is diagonal in the computational basis. Suppose that we apply the same
unitary to another Choi matrix $\rho _{\mathcal{A}_{q^{\prime }}}$, although
\begin{equation}
U_{q}\rho _{\mathcal{A}_{q^{\prime }}}U_{q}^{\dagger }=\left(
\begin{array}{cccc}
q_{00} & 0 & \frac{q\sqrt{1-q^{\prime }}-\sqrt{1-q}q^{\prime }}{4-2q} & 0 \\
0 & 0 & 0 & 0 \\
\frac{q\sqrt{1-q^{\prime }}-\sqrt{1-q}q^{\prime }}{4-2q} & 0 & 1-\frac{q^{\prime }}{2}-q_{00} & 0 \\
0 & 0 & 0 & \frac{q^{\prime }}{2}%
\end{array}%
\right),
\end{equation}%
is not diagonal, the diagonal part gives the probability distribution over
the computational basis, i.e., $\{q_{00},0,1-{q^{\prime }}/{2}-q_{00},{%
q^{\prime }}/{2}\}$, where $q_{00}={\left( 2-q-q^{\prime }-2\sqrt{%
(1-q)(1-q^{\prime })}\right) }/{\left( 4-2q\right) }$.

In a binary discrimination problem between $\rho _{\mathcal{A}_{q_{0}}}$ and
$\rho _{\mathcal{A}_{q_{1}}}$, one can simply apply $U_{q_{0}}^{\otimes u}$
or $U_{q_{1}}^{\otimes u}$ and then measure in the Bell basis. A direct
maximum-likelihood decision can be used for the final decision. The error
probability can be calculated numerically. It also turns out that applying $%
U_{\min (q_{0},q_{1})}$ gives a slightly better performance.

\ 

\

\end{widetext}


\begin{thebibliography}{45}%
\makeatletter
\providecommand \@ifxundefined [1]{%
 \@ifx{#1\undefined}
}%
\providecommand \@ifnum [1]{%
 \ifnum #1\expandafter \@firstoftwo
 \else \expandafter \@secondoftwo
 \fi
}%
\providecommand \@ifx [1]{%
 \ifx #1\expandafter \@firstoftwo
 \else \expandafter \@secondoftwo
 \fi
}%
\providecommand \natexlab [1]{#1}%
\providecommand \enquote  [1]{``#1''}%
\providecommand \bibnamefont  [1]{#1}%
\providecommand \bibfnamefont [1]{#1}%
\providecommand \citenamefont [1]{#1}%
\providecommand \href@noop [0]{\@secondoftwo}%
\providecommand \href [0]{\begingroup \@sanitize@url \@href}%
\providecommand \@href[1]{\@@startlink{#1}\@@href}%
\providecommand \@@href[1]{\endgroup#1\@@endlink}%
\providecommand \@sanitize@url [0]{\catcode `\\12\catcode `\$12\catcode
  `\&12\catcode `\#12\catcode `\^12\catcode `\_12\catcode `\%12\relax}%
\providecommand \@@startlink[1]{}%
\providecommand \@@endlink[0]{}%
\providecommand \url  [0]{\begingroup\@sanitize@url \@url }%
\providecommand \@url [1]{\endgroup\@href {#1}{\urlprefix }}%
\providecommand \urlprefix  [0]{URL }%
\providecommand \Eprint [0]{\href }%
\providecommand \doibase [0]{https://doi.org/}%
\providecommand \selectlanguage [0]{\@gobble}%
\providecommand \bibinfo  [0]{\@secondoftwo}%
\providecommand \bibfield  [0]{\@secondoftwo}%
\providecommand \translation [1]{[#1]}%
\providecommand \BibitemOpen [0]{}%
\providecommand \bibitemStop [0]{}%
\providecommand \bibitemNoStop [0]{.\EOS\space}%
\providecommand \EOS [0]{\spacefactor3000\relax}%
\providecommand \BibitemShut  [1]{\csname bibitem#1\endcsname}%
\let\auto@bib@innerbib\@empty
\bibitem [{\citenamefont {Helstrom}(1976)}]{Helstrom_1976}%
  \BibitemOpen
  \bibfield  {author} {\bibinfo {author} {\bibfnamefont {C.}~\bibnamefont
  {Helstrom}},\ }\href {https://books.google.com/books?id=fv9SAAAAMAAJ} {\emph
  {\bibinfo {title} {Quantum Detection and Estimation Theory}}},\ Mathematics
  in Science and Engineering: a series of monographs and textbooks\ (\bibinfo
  {publisher} {Academic Press},\ \bibinfo {year} {1976})\BibitemShut {NoStop}%
\bibitem [{\citenamefont {Chefles}\ and\ \citenamefont
  {Barnett}(1998)}]{Anthony_1998}%
  \BibitemOpen
  \bibfield  {author} {\bibinfo {author} {\bibfnamefont {A.}~\bibnamefont
  {Chefles}}\ and\ \bibinfo {author} {\bibfnamefont {S.~M.}\ \bibnamefont
  {Barnett}},\ }\bibfield  {title} {\bibinfo {title} {Quantum state separation,
  unambiguous discrimination and exact cloning},\ }\href@noop {} {\bibfield
  {journal} {\bibinfo  {journal} {J. Phys. A: Math. Gen.}\ }\textbf {\bibinfo
  {volume} {31}},\ \bibinfo {pages} {10097} (\bibinfo {year}
  {1998})}\BibitemShut {NoStop}%
\bibitem [{\citenamefont {Chefles}(2000)}]{Chefles_2000}%
  \BibitemOpen
  \bibfield  {author} {\bibinfo {author} {\bibfnamefont {A.}~\bibnamefont
  {Chefles}},\ }\bibfield  {title} {\bibinfo {title} {Quantum state
  discrimination},\ }\href {https://doi.org/10.1080/00107510010002599}
  {\bibfield  {journal} {\bibinfo  {journal} {Contemp. Phys.}\ }\textbf
  {\bibinfo {volume} {41}},\ \bibinfo {pages} {401} (\bibinfo {year}
  {2000})}\BibitemShut {NoStop}%
\bibitem{Janos2010}
J. A. Bergou, Discrimination of quantum states, Journal of Modern Optics \textbf{57}, 160-180 (2010).
\bibitem [{\citenamefont {Kitaev}(1997)}]{KitaevDiamond}%
  \BibitemOpen
  \bibfield  {author} {\bibinfo {author} {\bibfnamefont {A.~Y.}\ \bibnamefont
  {Kitaev}},\ }\bibfield  {title} {\bibinfo {title} {Quantum computations:
  algorithms and error correction},\ }\href
  {https://doi.org/10.1038/s41566-018-0301-6} {\bibfield  {journal} {\bibinfo
  {journal} {Russ. Math. Surv.}\ }\textbf {\bibinfo {volume} {52}},\ \bibinfo
  {pages} {1191} (\bibinfo {year} {1997})}\BibitemShut {NoStop}%
\bibitem [{\citenamefont {Ac\'{\i}n}\ \emph {et~al.}(2001)\citenamefont
  {Ac\'{\i}n}, \citenamefont {Jan\'e},\ and\ \citenamefont
  {Vidal}}]{Acin_2001}%
  \BibitemOpen
  \bibfield  {author} {\bibinfo {author} {\bibfnamefont {A.}~\bibnamefont
  {Ac\'{\i}n}}, \bibinfo {author} {\bibfnamefont {E.}~\bibnamefont {Jan\'e}},\
  and\ \bibinfo {author} {\bibfnamefont {G.}~\bibnamefont {Vidal}},\ }\bibfield
   {title} {\bibinfo {title} {Optimal estimation of quantum dynamics},\ }\href
  {https://doi.org/10.1103/PhysRevA.64.050302} {\bibfield  {journal} {\bibinfo
  {journal} {Phys. Rev. A}\ }\textbf {\bibinfo {volume} {64}},\ \bibinfo
  {pages} {050302(R)} (\bibinfo {year} {2001})}\BibitemShut {NoStop}%
\bibitem [{\citenamefont {Sacchi}(2005)}]{sacchi2005entanglement}%
  \BibitemOpen
  \bibfield  {author} {\bibinfo {author} {\bibfnamefont {M.~F.}\ \bibnamefont
  {Sacchi}},\ }\bibfield  {title} {\bibinfo {title} {Entanglement can enhance
  the distinguishability of entanglement-breaking channels},\ }\href@noop {}
  {\bibfield  {journal} {\bibinfo  {journal} {Phys. Rev. A}\ }\textbf {\bibinfo
  {volume} {72}},\ \bibinfo {pages} {014305} (\bibinfo {year}
  {2005})}\BibitemShut {NoStop}%
\bibitem [{\citenamefont {Wang}\ and\ \citenamefont
  {Ying}(2006)}]{wang2006unambiguous}%
  \BibitemOpen
  \bibfield  {author} {\bibinfo {author} {\bibfnamefont {G.}~\bibnamefont
  {Wang}}\ and\ \bibinfo {author} {\bibfnamefont {M.}~\bibnamefont {Ying}},\
  }\bibfield  {title} {\bibinfo {title} {Unambiguous discrimination among
  quantum operations},\ }\href@noop {} {\bibfield  {journal} {\bibinfo
  {journal} {Phys. Rev. A}\ }\textbf {\bibinfo {volume} {73}},\ \bibinfo
  {pages} {042301} (\bibinfo {year} {2006})}\BibitemShut {NoStop}%
\bibitem [{\citenamefont {Pirandola}\ \emph
  {et~al.}(2018{\natexlab{a}})\citenamefont {Pirandola}, \citenamefont
  {Bardhan}, \citenamefont {Gehring}, \citenamefont {Weedbrook},\ and\
  \citenamefont {Lloyd}}]{pirandola2018advances}%
  \BibitemOpen
  \bibfield  {author} {\bibinfo {author} {\bibfnamefont {S.}~\bibnamefont
  {Pirandola}}, \bibinfo {author} {\bibfnamefont {B.~R.}\ \bibnamefont
  {Bardhan}}, \bibinfo {author} {\bibfnamefont {T.}~\bibnamefont {Gehring}},
  \bibinfo {author} {\bibfnamefont {C.}~\bibnamefont {Weedbrook}},\ and\
  \bibinfo {author} {\bibfnamefont {S.}~\bibnamefont {Lloyd}},\ }\bibfield
  {title} {\bibinfo {title} {Advances in photonic quantum sensing},\
  }\href@noop {} {\bibfield  {journal} {\bibinfo  {journal} {Nat. Photonics}\
  }\textbf {\bibinfo {volume} {12}},\ \bibinfo {pages} {724} (\bibinfo {year}
  {2018}{\natexlab{a}})}\BibitemShut {NoStop}%
\bibitem [{\citenamefont {Tan}\ \emph {et~al.}(2008)\citenamefont {Tan},
  \citenamefont {Erkmen}, \citenamefont {Giovannetti}, \citenamefont {Guha},
  \citenamefont {Lloyd}, \citenamefont {Maccone}, \citenamefont {Pirandola},\
  and\ \citenamefont {Shapiro}}]{tan2008quantum}%
  \BibitemOpen
  \bibfield  {author} {\bibinfo {author} {\bibfnamefont {S.-H.}\ \bibnamefont
  {Tan}}, \bibinfo {author} {\bibfnamefont {B.~I.}\ \bibnamefont {Erkmen}},
  \bibinfo {author} {\bibfnamefont {V.}~\bibnamefont {Giovannetti}}, \bibinfo
  {author} {\bibfnamefont {S.}~\bibnamefont {Guha}}, \bibinfo {author}
  {\bibfnamefont {S.}~\bibnamefont {Lloyd}}, \bibinfo {author} {\bibfnamefont
  {L.}~\bibnamefont {Maccone}}, \bibinfo {author} {\bibfnamefont
  {S.}~\bibnamefont {Pirandola}},\ and\ \bibinfo {author} {\bibfnamefont
  {J.~H.}\ \bibnamefont {Shapiro}},\ }\bibfield  {title} {\bibinfo {title}
  {Quantum illumination with gaussian states},\ }\href@noop {} {\bibfield
  {journal} {\bibinfo  {journal} {Phys. Rev. Lett.}\ }\textbf {\bibinfo
  {volume} {101}},\ \bibinfo {pages} {253601} (\bibinfo {year}
  {2008})}\BibitemShut {NoStop}%
\bibitem [{\citenamefont {Zhuang}\ \emph
  {et~al.}(2017{\natexlab{a}})\citenamefont {Zhuang}, \citenamefont {Zhang},\
  and\ \citenamefont {Shapiro}}]{zhuang2017optimum}%
  \BibitemOpen
  \bibfield  {author} {\bibinfo {author} {\bibfnamefont {Q.}~\bibnamefont
  {Zhuang}}, \bibinfo {author} {\bibfnamefont {Z.}~\bibnamefont {Zhang}},\ and\
  \bibinfo {author} {\bibfnamefont {J.~H.}\ \bibnamefont {Shapiro}},\
  }\bibfield  {title} {\bibinfo {title} {Optimum mixed-state discrimination for
  noisy entanglement-enhanced sensing},\ }\href
  {https://doi.org/10.1103/PhysRevLett.118.040801} {\bibfield  {journal}
  {\bibinfo  {journal} {Phys. Rev. Lett.}\ }\textbf {\bibinfo {volume} {118}},\
  \bibinfo {pages} {040801} (\bibinfo {year} {2017}{\natexlab{a}})}\BibitemShut
  {NoStop}%
\bibitem [{\citenamefont {Zhuang}\ \emph
  {et~al.}(2017{\natexlab{b}})\citenamefont {Zhuang}, \citenamefont {Zhang},\
  and\ \citenamefont {Shapiro}}]{zhuang2017NP}%
  \BibitemOpen
  \bibfield  {author} {\bibinfo {author} {\bibfnamefont {Q.}~\bibnamefont
  {Zhuang}}, \bibinfo {author} {\bibfnamefont {Z.}~\bibnamefont {Zhang}},\ and\
  \bibinfo {author} {\bibfnamefont {J.~H.}\ \bibnamefont {Shapiro}},\
  }\bibfield  {title} {\bibinfo {title} {Entanglement-enhanced neyman--pearson
  target detection using quantum illumination},\ }\href@noop {} {\bibfield
  {journal} {\bibinfo  {journal} {JOSA B}\ }\textbf {\bibinfo {volume} {34}},\
  \bibinfo {pages} {1567} (\bibinfo {year} {2017}{\natexlab{b}})}\BibitemShut
  {NoStop}%
\bibitem [{\citenamefont {Zhuang}\ \emph
  {et~al.}(2017{\natexlab{c}})\citenamefont {Zhuang}, \citenamefont {Zhang},\
  and\ \citenamefont {Shapiro}}]{zhuang2017quantum}%
  \BibitemOpen
  \bibfield  {author} {\bibinfo {author} {\bibfnamefont {Q.}~\bibnamefont
  {Zhuang}}, \bibinfo {author} {\bibfnamefont {Z.}~\bibnamefont {Zhang}},\ and\
  \bibinfo {author} {\bibfnamefont {J.~H.}\ \bibnamefont {Shapiro}},\
  }\bibfield  {title} {\bibinfo {title} {Quantum illumination for enhanced
  detection of rayleigh-fading targets},\ }\href@noop {} {\bibfield  {journal}
  {\bibinfo  {journal} {Phys. Rev. A}\ }\textbf {\bibinfo {volume} {96}},\
  \bibinfo {pages} {020302(R)} (\bibinfo {year} {2017}{\natexlab{c}})}\BibitemShut
  {NoStop}%
\bibitem [{\citenamefont {Zhang}\ \emph {et~al.}(2015)\citenamefont {Zhang},
  \citenamefont {Mouradian}, \citenamefont {Wong},\ and\ \citenamefont
  {Shapiro}}]{zhang2015}%
  \BibitemOpen
  \bibfield  {author} {\bibinfo {author} {\bibfnamefont {Z.}~\bibnamefont
  {Zhang}}, \bibinfo {author} {\bibfnamefont {S.}~\bibnamefont {Mouradian}},
  \bibinfo {author} {\bibfnamefont {F.~N.~C.}\ \bibnamefont {Wong}},\ and\
  \bibinfo {author} {\bibfnamefont {J.~H.}\ \bibnamefont {Shapiro}},\
  }\bibfield  {title} {\bibinfo {title} {Entanglement-enhanced sensing in a
  lossy and noisy environment},\ }\href
  {https://doi.org/10.1103/PhysRevLett.114.110506} {\bibfield  {journal}
  {\bibinfo  {journal} {Phys. Rev. Lett.}\ }\textbf {\bibinfo {volume} {114}},\
  \bibinfo {pages} {110506} (\bibinfo {year} {2015})}\BibitemShut {NoStop}%
\bibitem [{\citenamefont {Pirandola}(2011)}]{Qreading}%
  \BibitemOpen
  \bibfield  {author} {\bibinfo {author} {\bibfnamefont {S.}~\bibnamefont
  {Pirandola}},\ }\bibfield  {title} {\bibinfo {title} {Quantum reading of a
  classical digital memory},\ }\href
  {https://doi.org/10.1103/PhysRevLett.106.090504} {\bibfield  {journal}
  {\bibinfo  {journal} {Phys. Rev. Lett.}\ }\textbf {\bibinfo {volume} {106}},\
  \bibinfo {pages} {090504} (\bibinfo {year} {2011})}\BibitemShut {NoStop}%
\bibitem [{\citenamefont {Takagi}\ \emph {et~al.}(2019)\citenamefont {Takagi},
  \citenamefont {Regula}, \citenamefont {Bu}, \citenamefont {Liu},\ and\
  \citenamefont {Adesso}}]{takagi2019operational}%
  \BibitemOpen
  \bibfield  {author} {\bibinfo {author} {\bibfnamefont {R.}~\bibnamefont
  {Takagi}}, \bibinfo {author} {\bibfnamefont {B.}~\bibnamefont {Regula}},
  \bibinfo {author} {\bibfnamefont {K.}~\bibnamefont {Bu}}, \bibinfo {author}
  {\bibfnamefont {Z.-W.}\ \bibnamefont {Liu}},\ and\ \bibinfo {author}
  {\bibfnamefont {G.}~\bibnamefont {Adesso}},\ }\bibfield  {title} {\bibinfo
  {title} {Operational advantage of quantum resources in subchannel
  discrimination},\ }\href@noop {} {\bibfield  {journal} {\bibinfo  {journal}
  {Phys. Rev. Lett.}\ }\textbf {\bibinfo {volume} {122}},\ \bibinfo {pages}
  {140402} (\bibinfo {year} {2019})}\BibitemShut {NoStop}%
\bibitem [{\citenamefont {Harrow}\ \emph {et~al.}(2010)\citenamefont {Harrow},
  \citenamefont {Hassidim}, \citenamefont {Leung},\ and\ \citenamefont
  {Watrous}}]{harrow2010adaptive}%
  \BibitemOpen
  \bibfield  {author} {\bibinfo {author} {\bibfnamefont {A.~W.}\ \bibnamefont
  {Harrow}}, \bibinfo {author} {\bibfnamefont {A.}~\bibnamefont {Hassidim}},
  \bibinfo {author} {\bibfnamefont {D.~W.}\ \bibnamefont {Leung}},\ and\
  \bibinfo {author} {\bibfnamefont {J.}~\bibnamefont {Watrous}},\ }\bibfield
  {title} {\bibinfo {title} {Adaptive versus nonadaptive strategies for quantum
  channel discrimination},\ }\href@noop {} {\bibfield  {journal} {\bibinfo
  {journal} {Phys. Rev. A}\ }\textbf {\bibinfo {volume} {81}},\ \bibinfo
  {pages} {032339} (\bibinfo {year} {2010})}\BibitemShut {NoStop}%
\bibitem [{\citenamefont {Acin}(2001)}]{acin2001statistical}%
  \BibitemOpen
  \bibfield  {author} {\bibinfo {author} {\bibfnamefont {A.}~\bibnamefont
  {Acin}},\ }\bibfield  {title} {\bibinfo {title} {Statistical
  distinguishability between unitary operations},\ }\href@noop {} {\bibfield
  {journal} {\bibinfo  {journal} {Phys. Rev. Lett.}\ }\textbf {\bibinfo
  {volume} {87}},\ \bibinfo {pages} {177901} (\bibinfo {year}
  {2001})}\BibitemShut {NoStop}%
\bibitem [{\citenamefont {Duan}\ \emph {et~al.}(2009)\citenamefont {Duan},
  \citenamefont {Feng},\ and\ \citenamefont {Ying}}]{duan2009perfect}%
  \BibitemOpen
  \bibfield  {author} {\bibinfo {author} {\bibfnamefont {R.}~\bibnamefont
  {Duan}}, \bibinfo {author} {\bibfnamefont {Y.}~\bibnamefont {Feng}},\ and\
  \bibinfo {author} {\bibfnamefont {M.}~\bibnamefont {Ying}},\ }\bibfield
  {title} {\bibinfo {title} {Perfect distinguishability of quantum
  operations},\ }\href {https://doi.org/10.1103/PhysRevLett.103.210501}
  {\bibfield  {journal} {\bibinfo  {journal} {Phys. Rev. Lett.}\ }\textbf
  {\bibinfo {volume} {103}},\ \bibinfo {pages} {210501} (\bibinfo {year}
  {2009})}\BibitemShut {NoStop}%
\bibitem [{\citenamefont {Duan}\ \emph {et~al.}(2007)\citenamefont {Duan},
  \citenamefont {Feng},\ and\ \citenamefont {Ying}}]{duan2007entanglement}%
  \BibitemOpen
  \bibfield  {author} {\bibinfo {author} {\bibfnamefont {R.}~\bibnamefont
  {Duan}}, \bibinfo {author} {\bibfnamefont {Y.}~\bibnamefont {Feng}},\ and\
  \bibinfo {author} {\bibfnamefont {M.}~\bibnamefont {Ying}},\ }\bibfield
  {title} {\bibinfo {title} {Entanglement is not necessary for perfect
  discrimination between unitary operations},\ }\href@noop {} {\bibfield
  {journal} {\bibinfo  {journal} {Phys. Rev. Lett.}\ }\textbf {\bibinfo
  {volume} {98}},\ \bibinfo {pages} {100503} (\bibinfo {year}
  {2007})}\BibitemShut {NoStop}%
\bibitem [{\citenamefont {Pirandola}\ \emph {et~al.}(2019)\citenamefont
  {Pirandola}, \citenamefont {Laurenza}, \citenamefont {Lupo},\ and\
  \citenamefont {Pereira}}]{pirandola2019fundamental}%
  \BibitemOpen
  \bibfield  {author} {\bibinfo {author} {\bibfnamefont {S.}~\bibnamefont
  {Pirandola}}, \bibinfo {author} {\bibfnamefont {R.}~\bibnamefont {Laurenza}},
  \bibinfo {author} {\bibfnamefont {C.}~\bibnamefont {Lupo}},\ and\ \bibinfo
  {author} {\bibfnamefont {J.~L.}\ \bibnamefont {Pereira}},\ }\bibfield
  {title} {\bibinfo {title} {Fundamental limits to quantum channel
  discrimination},\ }\href@noop {} {\bibfield  {journal} {\bibinfo  {journal}
  {Npj Quantum Inf.}\ }\textbf {\bibinfo {volume} {5}},\ \bibinfo {pages} {50}
  (\bibinfo {year} {2019})}\BibitemShut {NoStop}%
\bibitem [{\citenamefont {Pirandola}\ and\ \citenamefont
  {Lupo}(2017)}]{pirandola2017ultimate}%
  \BibitemOpen
  \bibfield  {author} {\bibinfo {author} {\bibfnamefont {S.}~\bibnamefont
  {Pirandola}}\ and\ \bibinfo {author} {\bibfnamefont {C.}~\bibnamefont
  {Lupo}},\ }\bibfield  {title} {\bibinfo {title} {Ultimate precision of
  adaptive noise estimation},\ }\href@noop {} {\bibfield  {journal} {\bibinfo
  {journal} {Phys. Rev. Lett.}\ }\textbf {\bibinfo {volume} {118}},\ \bibinfo
  {pages} {100502} (\bibinfo {year} {2017})}\BibitemShut {NoStop}%
\bibitem [{\citenamefont {Cariolaro}\ and\ \citenamefont
  {Pierobon}(2010)}]{cariolaro2010theory}%
  \BibitemOpen
  \bibfield  {author} {\bibinfo {author} {\bibfnamefont {G.}~\bibnamefont
  {Cariolaro}}\ and\ \bibinfo {author} {\bibfnamefont {G.}~\bibnamefont
  {Pierobon}},\ }\bibfield  {title} {\bibinfo {title} {Theory of quantum pulse
  position modulation and related numerical problems},\ }\href@noop {}
  {\bibfield  {journal} {\bibinfo  {journal} {IEEE Trans. Commun.}\ }\textbf
  {\bibinfo {volume} {58}},\ \bibinfo {pages} {1213} (\bibinfo {year}
  {2010})}\BibitemShut {NoStop}%
\bibitem{CPF2020}
\QZ{Q. Zhuang and S. Pirandola, Entanglement-enhanced testing of multiple quantum hypotheses, Commun. Phys. {\bf 3}, \SP{103} (2020).
}
\bibitem [{\citenamefont {Holevo}(1978)}]{PGM1}%
  \BibitemOpen
  \bibfield  {author} {\bibinfo {author} {\bibfnamefont {A.~S.}\ \bibnamefont
  {Holevo}},\ }\bibfield  {title} {\bibinfo {title} {On asymptotically optimal
  hypotheses testing in quantum statistics},\ }\href@noop {} {\bibfield
  {journal} {\bibinfo  {journal} {Teoriya Veroyatnostei i ee Primeneniya}\
  }\textbf {\bibinfo {volume} {23}},\ \bibinfo {pages} {429} (\bibinfo {year}
  {1978})}\BibitemShut {NoStop}%
\bibitem [{\citenamefont {Hausladen}\ and\ \citenamefont
  {Wootters}(1994)}]{PGM2}%
  \BibitemOpen
  \bibfield  {author} {\bibinfo {author} {\bibfnamefont {P.}~\bibnamefont
  {Hausladen}}\ and\ \bibinfo {author} {\bibfnamefont {W.~K.}\ \bibnamefont
  {Wootters}},\ }\bibfield  {title} {\bibinfo {title} {A `pretty good'
  measurement for distinguishing quantum states},\ }\href@noop {} {\bibfield
  {journal} {\bibinfo  {journal} {J. Mod. Opt.}\ }\textbf {\bibinfo {volume}
  {41}},\ \bibinfo {pages} {2385} (\bibinfo {year} {1994})}\BibitemShut
  {NoStop}%
\bibitem [{\citenamefont {Hausladen}\ \emph {et~al.}(1996)\citenamefont
  {Hausladen}, \citenamefont {Jozsa}, \citenamefont {Schumacher}, \citenamefont
  {Westmoreland},\ and\ \citenamefont {Wootters}}]{PGM3}%
  \BibitemOpen
  \bibfield  {author} {\bibinfo {author} {\bibfnamefont {P.}~\bibnamefont
  {Hausladen}}, \bibinfo {author} {\bibfnamefont {R.}~\bibnamefont {Jozsa}},
  \bibinfo {author} {\bibfnamefont {B.}~\bibnamefont {Schumacher}}, \bibinfo
  {author} {\bibfnamefont {M.}~\bibnamefont {Westmoreland}},\ and\ \bibinfo
  {author} {\bibfnamefont {W.~K.}\ \bibnamefont {Wootters}},\ }\bibfield
  {title} {\bibinfo {title} {Classical information capacity of a quantum
  channel},\ }\href {https://doi.org/10.1103/PhysRevA.54.1869} {\bibfield
  {journal} {\bibinfo  {journal} {Phys. Rev. A}\ }\textbf {\bibinfo {volume}
  {54}},\ \bibinfo {pages} {1869} (\bibinfo {year} {1996})}\BibitemShut
  {NoStop}%
\bibitem [{\citenamefont {Barnum}\ and\ \citenamefont {Knill}(2002)}]{Barnum}%
  \BibitemOpen
  \bibfield  {author} {\bibinfo {author} {\bibfnamefont {H.}~\bibnamefont
  {Barnum}}\ and\ \bibinfo {author} {\bibfnamefont {E.}~\bibnamefont {Knill}},\
  }\bibfield  {title} {\bibinfo {title} {Reversing quantum dynamics with
  near-optimal quantum and classical fidelity},\ }\href@noop {} {\bibfield
  {journal} {\bibinfo  {journal} {J. Math. Phys.}\ }\textbf {\bibinfo {volume}
  {43}},\ \bibinfo {pages} {2097} (\bibinfo {year} {2002})}\BibitemShut
  {NoStop}%
\bibitem [{\citenamefont {Bagan}\ \emph {et~al.}(2016)\citenamefont {Bagan},
  \citenamefont {Bergou}, \citenamefont {Cottrell},\ and\ \citenamefont
  {Hillery}}]{Bagan}%
  \BibitemOpen
  \bibfield  {author} {\bibinfo {author} {\bibfnamefont {E.}~\bibnamefont
  {Bagan}}, \bibinfo {author} {\bibfnamefont {J.~A.}\ \bibnamefont {Bergou}},
  \bibinfo {author} {\bibfnamefont {S.~S.}\ \bibnamefont {Cottrell}},\ and\
  \bibinfo {author} {\bibfnamefont {M.}~\bibnamefont {Hillery}},\ }\bibfield
  {title} {\bibinfo {title} {Relations between coherence and path
  information},\ }\href@noop {} {\bibfield  {journal} {\bibinfo  {journal}
  {Phys. Rev. Lett.}\ }\textbf {\bibinfo {volume} {116}},\ \bibinfo {pages}
  {160406} (\bibinfo {year} {2016})}\BibitemShut {NoStop}%
\bibitem [{\citenamefont {Qiu}\ and\ \citenamefont {Li}(2010)}]{Qiu}%
  \BibitemOpen
  \bibfield  {author} {\bibinfo {author} {\bibfnamefont {D.}~\bibnamefont
  {Qiu}}\ and\ \bibinfo {author} {\bibfnamefont {L.}~\bibnamefont {Li}},\
  }\bibfield  {title} {\bibinfo {title} {Minimum-error discrimination of
  quantum states: Bounds and comparisons},\ }\href@noop {} {\bibfield
  {journal} {\bibinfo  {journal} {Phys. Rev. A}\ }\textbf {\bibinfo {volume}
  {81}},\ \bibinfo {pages} {042329} (\bibinfo {year} {2010})}\BibitemShut
  {NoStop}%
\bibitem [{\citenamefont {Ogawa}\ and\ \citenamefont {Nagaoka}(1999)}]{Ogawa}%
  \BibitemOpen
  \bibfield  {author} {\bibinfo {author} {\bibfnamefont {T.}~\bibnamefont
  {Ogawa}}\ and\ \bibinfo {author} {\bibfnamefont {H.}~\bibnamefont
  {Nagaoka}},\ }\bibfield  {title} {\bibinfo {title} {Strong converse to the
  quantum channel coding theorem},\ }\href@noop {} {\bibfield  {journal}
  {\bibinfo  {journal} {IEEE Trans. Inf. Theory}\ }\textbf {\bibinfo {volume}
  {45}},\ \bibinfo {pages} {2486} (\bibinfo {year} {1999})}\BibitemShut
  {NoStop}%
\bibitem [{\citenamefont {Montanaro}(2008)}]{montanaro2008lower}%
  \BibitemOpen
  \bibfield  {author} {\bibinfo {author} {\bibfnamefont {A.}~\bibnamefont
  {Montanaro}},\ }\bibfield  {title} {\bibinfo {title} {A lower bound on the
  probability of error in quantum state discrimination},\ }in\ \href@noop {}
  {\emph {\bibinfo {booktitle} {2008 IEEE Information Theory Workshop}}}\
  (\bibinfo {organization} {IEEE},\ \bibinfo {year} {2008})\ pp.\ \bibinfo
  {pages} {378--380}\BibitemShut {NoStop}%
\bibitem [{sup()}]{supp}%
See Supplemental Material [url] for details of proofs and calculations, which includes Refs.~\cite{FuchsGraaf,yuen1975optimum,eldar2004optimal}
\bibitem [{\citenamefont {Fuchs}\ and\ \citenamefont {van~de
  Graaf}(1999)}]{FuchsGraaf}%
  \BibitemOpen
  \bibfield  {author} {\bibinfo {author} {\bibfnamefont {C.~A.}\ \bibnamefont
  {Fuchs}}\ and\ \bibinfo {author} {\bibfnamefont {J.}~\bibnamefont {van~de
  Graaf}},\ }\bibfield  {title} {\bibinfo {title} {Cryptographic
  distinguishability measures for quantum mechanical states},\ }\href@noop {}
  {\bibfield  {journal} {\bibinfo  {journal} {IEEE Transactions on Information
  Theory}\ }\textbf {\bibinfo {volume} {45}},\ \bibinfo {pages} {1216}
  (\bibinfo {year} {1999})}\BibitemShut {NoStop}%
\bibitem [{\citenamefont {Yuen}\ \emph {et~al.}(1975)\citenamefont {Yuen},
  \citenamefont {Kennedy},\ and\ \citenamefont {Lax}}]{yuen1975optimum}%
  \BibitemOpen
  \bibfield  {author} {\bibinfo {author} {\bibfnamefont {H.}~\bibnamefont
  {Yuen}}, \bibinfo {author} {\bibfnamefont {R.}~\bibnamefont {Kennedy}},\ and\
  \bibinfo {author} {\bibfnamefont {M.}~\bibnamefont {Lax}},\ }\bibfield
  {title} {\bibinfo {title} {Optimum testing of multiple hypotheses in quantum
  detection theory},\ }\href@noop {} {\bibfield  {journal} {\bibinfo  {journal}
  {IEEE Trans. Inf. Theory}\ }\textbf {\bibinfo {volume} {21}},\ \bibinfo
  {pages} {125} (\bibinfo {year} {1975})}\BibitemShut {NoStop}%
\bibitem [{\citenamefont {Eldar}\ \emph {et~al.}(2004)\citenamefont {Eldar},
  \citenamefont {Megretski},\ and\ \citenamefont
  {Verghese}}]{eldar2004optimal}%
  \BibitemOpen
  \bibfield  {author} {\bibinfo {author} {\bibfnamefont {Y.~C.}\ \bibnamefont
  {Eldar}}, \bibinfo {author} {\bibfnamefont {A.}~\bibnamefont {Megretski}},\
  and\ \bibinfo {author} {\bibfnamefont {G.~C.}\ \bibnamefont {Verghese}},\
  }\bibfield  {title} {\bibinfo {title} {Optimal detection of symmetric mixed
  quantum states},\ }\href@noop {} {\bibfield  {journal} {\bibinfo  {journal}
  {IEEE Trans. Inf. Theory}\ }\textbf {\bibinfo {volume} {50}},\ \bibinfo
  {pages} {1198} (\bibinfo {year} {2004})}\BibitemShut {NoStop}%
\bibitem [{\citenamefont {Nielsen}\ and\ \citenamefont
  {Chuang}(1997)}]{quantumPQGA}%
  \BibitemOpen
  \bibfield  {author} {\bibinfo {author} {\bibfnamefont {M.~A.}\ \bibnamefont
  {Nielsen}}\ and\ \bibinfo {author} {\bibfnamefont {I.~L.}\ \bibnamefont
  {Chuang}},\ }\bibfield  {title} {\bibinfo {title} {Programmable quantum gate
  arrays},\ }\href@noop {} {\bibfield  {journal} {\bibinfo  {journal} {Phys.
  Rev. Lett.}\ }\textbf {\bibinfo {volume} {79}},\ \bibinfo {pages} {321}
  (\bibinfo {year} {1997})}\BibitemShut {NoStop}%
\bibitem [{\citenamefont {Pirandola}\ \emph {et~al.}(2017)\citenamefont
  {Pirandola}, \citenamefont {Laurenza}, \citenamefont {Ottaviani},\ and\
  \citenamefont {Banchi}}]{PLOB}%
  \BibitemOpen
  \bibfield  {author} {\bibinfo {author} {\bibfnamefont {S.}~\bibnamefont
  {Pirandola}}, \bibinfo {author} {\bibfnamefont {R.}~\bibnamefont {Laurenza}},
  \bibinfo {author} {\bibfnamefont {C.}~\bibnamefont {Ottaviani}},\ and\
  \bibinfo {author} {\bibfnamefont {L.}~\bibnamefont {Banchi}},\ }\bibfield
  {title} {\bibinfo {title} {Fundamental limits of repeaterless quantum
  communications},\ }\href@noop {} {\bibfield  {journal} {\bibinfo  {journal}
  {Nat. Commun.}\ }\textbf {\bibinfo {volume} {8}},\ \bibinfo {pages} {15043}
  (\bibinfo {year} {2017})}\BibitemShut {NoStop}%
\SP{\bibitem {TQCtheory}
  \BibitemOpen
  \bibfield  {author} {\bibinfo {author} {\bibfnamefont {L.}~\bibnamefont
  {Banchi}}, \bibinfo {author} {\bibfnamefont {J.}~\bibnamefont
  {Pereira}}, \bibinfo {author} {\bibfnamefont {S.}~\bibnamefont
  {Lloyd}},\ and\ \bibinfo {author} {\bibfnamefont {S.}~\bibnamefont {Pirandola}},\ }\bibfield  {title}
  {\bibinfo {title} {Convex optimization of programmable quantum computers},\ }\href@noop {}
  {\bibfield  {journal} {\bibinfo  {journal} {npj Quantum Information}\ }\textbf
  {\bibinfo {volume} {6}},\ \bibinfo {pages} {42} (\bibinfo {year}
  {2020}{\natexlab{b}})}\BibitemShut {NoStop}}%
\bibitem [{\citenamefont {Ishizaka}\ and\ \citenamefont
  {Hiroshima}(2008)}]{ishizaka2008asymptotic}%
  \BibitemOpen
  \bibfield  {author} {\bibinfo {author} {\bibfnamefont {S.}~\bibnamefont
  {Ishizaka}}\ and\ \bibinfo {author} {\bibfnamefont {T.}~\bibnamefont
  {Hiroshima}},\ }\bibfield  {title} {\bibinfo {title} {Asymptotic
  teleportation scheme as a universal programmable quantum processor},\
  }\href@noop {} {\bibfield  {journal} {\bibinfo  {journal} {Phys. Rev. Lett.}\
  }\textbf {\bibinfo {volume} {101}},\ \bibinfo {pages} {240501} (\bibinfo
  {year} {2008})}\BibitemShut {NoStop}%
\bibitem [{\citenamefont {Paulsen}(2002)}]{PaulsenBook}%
  \BibitemOpen
  \bibfield  {author} {\bibinfo {author} {\bibfnamefont {V.~I.}\ \bibnamefont
  {Paulsen}},\ }\href@noop {} {\emph {\bibinfo {title} {Completely Bounded Maps
  and Operator Algebras}}}\ (\bibinfo  {publisher} {Cambridge University
  Press},\ \bibinfo {year} {2002})\BibitemShut {NoStop}%
\bibitem [{foo({\natexlab{a}})}]{footnote1}%
In the following, where we refer to a
  $d$-dimensional channel, we refer to the dimension of the input Hilbert space. the dimension of the output Hilbert space could be different, as it
  happens, for instance, for an erasure channel.
\bibitem [{\citenamefont {Holevo}(2002)}]{holevo2002remarks}%
  \BibitemOpen
  \bibfield  {author} {\bibinfo {author} {\bibfnamefont {A.~S.}\ \bibnamefont
  {Holevo}},\ }\bibfield  {title} {\bibinfo {title} {Remarks on the classical
  capacity of quantum channel},\ }\href@noop {} {\bibfield  {journal} {\bibinfo
   {journal} {arXiv 0212025}\ } (\bibinfo {year} {2002})}\BibitemShut {NoStop}%
\bibitem [{\citenamefont {Datta}\ \emph {et~al.}(2006)\citenamefont {Datta},
  \citenamefont {Fukuda},\ and\ \citenamefont
  {Holevo}}]{datta2006complementarity}%
  \BibitemOpen
  \bibfield  {author} {\bibinfo {author} {\bibfnamefont {N.}~\bibnamefont
  {Datta}}, \bibinfo {author} {\bibfnamefont {M.}~\bibnamefont {Fukuda}},\ and\
  \bibinfo {author} {\bibfnamefont {A.~S.}\ \bibnamefont {Holevo}},\ }\bibfield
   {title} {\bibinfo {title} {Complementarity and additivity for covariant
  channels},\ }\href@noop {} {\bibfield  {journal} {\bibinfo  {journal}
  {Quantum Inf. Process.}\ }\textbf {\bibinfo {volume} {5}},\ \bibinfo {pages}
  {179} (\bibinfo {year} {2006})}\BibitemShut {NoStop}%
\bibitem [{\citenamefont {Zhuang}\ \emph
  {et~al.}(2017{\natexlab{d}})\citenamefont {Zhuang}, \citenamefont {Zhu},\
  and\ \citenamefont {Shor}}]{zhuang2017additive}%
  \BibitemOpen
  \bibfield  {author} {\bibinfo {author} {\bibfnamefont {Q.}~\bibnamefont
  {Zhuang}}, \bibinfo {author} {\bibfnamefont {E.~Y.}\ \bibnamefont {Zhu}},\
  and\ \bibinfo {author} {\bibfnamefont {P.~W.}\ \bibnamefont {Shor}},\
  }\bibfield  {title} {\bibinfo {title} {Additive classical capacity of quantum
  channels assisted by noisy entanglement},\ }\href@noop {} {\bibfield
  {journal} {\bibinfo  {journal} {Phys. Rev. Lett.}\ }\textbf {\bibinfo
  {volume} {118}},\ \bibinfo {pages} {200503} (\bibinfo {year}
  {2017}{\natexlab{d}})}\BibitemShut {NoStop}%
\bibitem [{\citenamefont {Bennett}\ and\ \citenamefont
  {Wiesner}(1992)}]{bennett1992}%
  \BibitemOpen
  \bibfield  {author} {\bibinfo {author} {\bibfnamefont {C.~H.}\ \bibnamefont
  {Bennett}}\ and\ \bibinfo {author} {\bibfnamefont {S.~J.}\ \bibnamefont
  {Wiesner}},\ }\bibfield  {title} {\bibinfo {title} {Communication via one-and
  two-particle operators on einstein-podolsky-rosen states},\ }\href@noop {}
  {\bibfield  {journal} {\bibinfo  {journal} {Phys. Rev. Lett.}\ }\textbf
  {\bibinfo {volume} {69}},\ \bibinfo {pages} {2881} (\bibinfo {year}
  {1992})}\BibitemShut {NoStop}%
\bibitem [{\citenamefont {Dalla~Pozza}\ and\ \citenamefont
  {Pierobon}(2015)}]{dalla2015optimality}%
  \BibitemOpen
  \bibfield  {author} {\bibinfo {author} {\bibfnamefont {N.}~\bibnamefont
  {Dalla~Pozza}}\ and\ \bibinfo {author} {\bibfnamefont {G.}~\bibnamefont
  {Pierobon}},\ }\bibfield  {title} {\bibinfo {title} {Optimality of
  square-root measurements in quantum state discrimination},\ }\href@noop {}
  {\bibfield  {journal} {\bibinfo  {journal} {Phys. Rev. A}\ }\textbf {\bibinfo
  {volume} {91}},\ \bibinfo {pages} {042334} (\bibinfo {year}
  {2015})}\BibitemShut {NoStop}%
\bibitem [{foo({\natexlab{b}})}]{footnote2}%
This comes from Eq.~(\ref{bound1_F}) where
  we use the expression of the multi-channel Choi matrix in
  Eq.~(\ref{TMSV_out}), the multiplicativity of the fidelity under tensor
  products, and the fact that the priors are equal, so that
  $\sum_{k^\prime>k}p_{k^\prime}p_{k} \rightarrow \frac{m-1}{2m}$. The
  expression of the simulation error can also be exploited in
  Eq.~(\ref{LB_GUS}) which explicitly accounts for the GUS property of the CPF
  problem.
\bibitem [{foo({\natexlab{c}})}]{footnote3}%
Here $\vert \bm x \vert=\sum_i |\bm x_i|$ is
  the vector one-norm.
\bibitem{Janos2002}
Y. Sun, J. A. Bergou, and M. Hillery, Optimum unambiguous discrimination between subsets of nonorthogonal quantum states, Phys. Rev. A \textbf{66}, 032315 (2002).
\bibitem{USD1}
A. Chefles, Condition for unambiguous state discrimination using local operations and classical communication, Phys. Rev. A {\bf 69}, 050307(R) (2004).
\bibitem{Janos2005}
U. Herzog, and J. A. Bergou, Optimum unambiguous discrimination of two mixed quantum states,
Phys. Rev. A 71, 050301 (2005).
\bibitem{USD2} 
M. Kleinmann, H. Kampermann, and D. Bru\ss, Unambiguous discrimination of mixed quantum states: Optimal solution and case study, Phys. Rev. A {\bf 81}, 020304(R) (2010).
\end{thebibliography}

\end{document}